\documentclass[onecolumn,secnumarabic,amsmath,amssymb,balancelastpage,nofootinbib]{article}

\usepackage[e]{esvect}
\usepackage{color}         
\usepackage{graphics}      
\usepackage{graphicx}      
\usepackage{epsf}          
\usepackage{bm}            

\usepackage{sectsty}
\allsectionsfont{\centering}  
                                       
\usepackage{natbib}
\usepackage{amssymb}
\usepackage{amsmath}
\usepackage{mathrsfs}
\usepackage{framed}
\usepackage{bigints}
\usepackage{enumitem}
\usepackage{setspace}
\usepackage{yfonts}

\usepackage[colorlinks=true]{hyperref}  

\newcommand{\mockalph}[1]{} 

\definecolor{darkred}{rgb}{0.6,0,0}
\definecolor{darkgreen}{rgb}{0,0.5,0}
\definecolor{darkblue}{rgb}{0,0,0.6}
\hypersetup{ colorlinks,
linkcolor=darkblue,
filecolor=darkgreen,
urlcolor=darkgreen,
citecolor=darkred }

\newcommand{\bra}[1]{\ensuremath{\left\langle#1\right|}}
\newcommand{\ket}[1]{\ensuremath{\left|#1\right\rangle}}
\newcommand{\bracket}[2]{\ensuremath{\left\langle #1 \middle| #2 \right\rangle}}

\newcommand{\calo}{{\mathcal O}}

\begin{document}
\bibliographystyle{authordate1}



\title{Self-Locating Uncertainty and the Origin of Probability in\vspace*{-.1cm}\\Everettian Quantum Mechanics\vspace*{-.3cm}}
\author         {\small{[arXiv v.3]}\vspace*{-.1cm}\\\small{CALT 68-2928}\vspace{0.25cm}\\Charles T. Sebens and Sean M. Carroll\vspace*{0.2cm}\\Forthcoming in \emph{The British Journal for the Philosophy of Science}\vspace{0.1cm}}
\date{May 29, 2015}

\maketitle
\begin{abstract}
A longstanding issue in attempts to understand the Everett (Many-Worlds) approach to quantum mechanics is the origin of the Born rule: why is the probability given by the square of the amplitude? Following Vaidman, we note that observers are in a position of self-locating uncertainty during the period between the branches of the wave function splitting via decoherence and the observer registering the outcome of the measurement. In this period it is tempting to regard each branch as equiprobable, but we argue that the temptation should be resisted.  Applying lessons from this analysis, we demonstrate (using methods similar to those of Zurek's envariance-based derivation) that the Born rule is the uniquely rational way of apportioning credence in Everettian quantum mechanics.  In doing so, we rely on a single key principle: changes purely to the environment do not affect the probabilities one ought to assign to measurement outcomes in a local subsystem.  We arrive at a method for assigning probabilities in cases that involve both classical and quantum self-locating uncertainty. This method provides unique answers to quantum Sleeping Beauty problems, as well as a well-defined procedure for calculating probabilities in quantum cosmological multiverses with multiple similar observers.
\end{abstract}
\defcitealias{everett}{Everett's} 
\defcitealias{albert2010}{Albert's} 
\defcitealias{wallace2012}{Wallace's}
\defcitealias{zurek2005}{Zurek's}
\defcitealias{zurek2003}{Zurek's}   
\newpage

\tableofcontents

\newpage

\section{Introduction}

The Everett (or Many-Worlds) approach to quantum mechanics is distinguished by its simplicity. The dynamics of the theory consists of a single deterministic evolution law, the Schr\"odinger equation.   There are no separate rules for dealing with `wave function collapse' and `quantum measurement,' nor are there additional hidden variables. A longstanding challenge for such an approach is to reproduce the Born rule: the probability of a measurement yielding eigenvalue $a$ of the observable $\widehat{A}$, given that the system was prepared in state $|\psi\rangle$, is $|\langle a|\psi\rangle|^2$, where $|a\rangle$ is the\footnote{Assuming there is a unique eigenstate with eigenvalue $a$.} eigenstate with eigenvalue $a$. Indeed, it seems like quite a challenge to explain how Everettian quantum mechanics could provide a theory of probabilities at all, given that states can be specified with arbitrary precision and all evolution is perfectly deterministic.

In this paper we argue that probability arises in Everettian quantum mechanics because observers with perfect knowledge (of their immediate circumstances and the state of the universe as a whole) necessarily evolve into conditions of self-locating uncertainty, in which they do not know which approximately isolated semi-classical world (or `branch') they inhabit.\footnote{The increase in the number of branches over time is a consequence of unitary evolution, not an additional postulate \citep{wallace2003, wallace2010c}.}  We propose a general principle, the Epistemic Separability Principle (\emph{ESP}), which captures the idea that predictions made by local agents should be independent of their environment. Given \emph{ESP}, we show that there is a unique rational way for such an agent to assign a credence to each of the quantum worlds they might be in.  These credences are precisely the ones recommended by the Born rule.  The probabilities are fundamentally \emph{subjective} in the sense that that they are not written into the laws---as they are in spontaneous collapse theories---but instead capture the degrees of belief of a rational agent; however, they are \emph{objective} in the sense that a rational agent  \emph{must} assign Born rule probabilities (if \emph{ESP} is correct).

We believe that recent proofs of the Born rule \citep{deutsch1999, wallace2010b, zurek2005} are on the right track, latching onto symmetries in the theory that serve to explain the validity of the Born rule in Everettian quantum mechanics.  However, there are serious objections to the approaches already explored and many remain unconvinced, so we offer this derivation as a novel alternative.  We seek to provide an \emph{epistemic}---as opposed to a \emph{decision-theoretic}---derivation of the Born rule which connects quantum uncertainty to the sort of self-locating uncertainty present in very large universes (discussed recently in \citealp{hartle2007, page2007, srednicki2010}). This approach shares formal features with \citetalias{zurek2003} \citeyearpar{zurek2003,zurek2003b,zurek2005} argument based on the idea of envariance, while offering a clearer explanation of the way in which probabilities arise in a deterministic setting.
 
At first glance, standard treatments of self-locating uncertainty seem to generate a serious problem for the many-worlds interpretation.  \citet{elga2004} has put forward a compelling principle of indifference for cases of self-locating uncertainty, roughly: an observer should give equal credence to any one of a discrete set of locations in the universe that are consistent with the data she has. A simplistic extension to the quantum case would seem to favor treating each world as equiprobable (branch-counting) rather than the Born rule, since Everettian observers on different branches find themselves in situations of identical data. We believe that the reasoning behind Elga's principle, when properly applied to Everettian quantum mechanics, actually leads to the Born rule---not branch-counting.  

One common assumption appealed to in determining what probabilities one ought to assign is that certain features of the case are irrelevant, in particular, one can alter the environment in many ways without changing what probability assignments are rational.  We formalize this idea by proposing a constraint on rational credences (\emph{ESP}) and showing that, somewhat surprisingly, the principle suffices to derive the Born rule.  A careful derivation will be provided after extensive stage-setting in \textsection \ref{prf} and appendix \ref{generalization}, but the spirit of the approach can be seen in this simplified version of the argument for two equal amplitude branches.  Imagine that there are two quantum measurement devices which measure the $z$-spin of an $x$-spin up particle, the first displaying either $\heartsuit$ or $\diamondsuit$, the second either $\clubsuit$ or $\spadesuit$ depending on the result of the measurement.  The post-measurement state of the detectors can be written as
\begin{equation}
\frac{1}{\sqrt{2}}\ket{\heartsuit}\ket{\clubsuit}+\frac{1}{\sqrt{2}}\ket{\diamondsuit}\ket{\spadesuit}\ .
\label{casenumberone}
\end{equation}
If we assume there is an experimenter who has yet to observe the outcome, they ought to be uncertain about which of the two branches they are on.  The probabilities assigned to the possible states of detector 1 ($\heartsuit$/$\diamondsuit$) should be the same if the case is modified so that detector 2 is wired to show the opposite symbols from those in \eqref{casenumberone} as the state of the second detector is irrelevant to to the experimenter's question about their relation to the first detector,
\begin{equation}
\frac{1}{\sqrt{2}}\ket{\heartsuit}\ket{\spadesuit}+\frac{1}{\sqrt{2}}\ket{\diamondsuit}\ket{\clubsuit}\ .
\label{casenumbertwo}
\end{equation}
The probability of $\heartsuit$ in \eqref{casenumberone}---which is the probability of being in the \emph{first} branch of \eqref{casenumberone}---is the same as that of $\heartsuit$ in \eqref{casenumbertwo}---which is the probability of being in the \emph{first} branch of \eqref{casenumbertwo}.  If instead we focus on assigning probabilities to the possible states of detector 2 ($\clubsuit$/$\spadesuit$), they too should be the same in \eqref{casenumberone} and \eqref{casenumbertwo}.  Focusing on $\spadesuit$ yields the requirement that the probability of being in the \emph{second} branch of \eqref{casenumberone} be the same as the probability of being in the \emph{first} branch of \eqref{casenumbertwo}.  And thus the probabilities assigned to each branch of \eqref{casenumberone} must be equal.

In \textsection \ref{prelim}, we begin by introducing the many-worlds interpretation and discussing how self-locating uncertainty arises during quantum measurements.  If we extend standard methods of handling self-locating uncertainty to quantum cases in a simple way, we get the odd recommendation that branches should be weighted equally regardless of their quantum amplitudes.  We argue that this cannot be the right way to handle the uncertainty that arises in cases of quantum branching.  In \textsection \ref{newf} we provide an alternative principle for handling self-locating uncertainty across physical theories (\emph{ESP}) and show that it yields the Born rule when applied to Everettian quantum mechanics.  In \textsection \ref{varieties}, we show that this principle (when strengthened) yields the result that each copy of oneself should be judged equiprobable in cases of classical self-locating uncertainty.  We then use the principle to generate probabilities in cases which involve both uncertainty arising from quantum measurement and from duplication of one's experiences elsewhere in the universe, such as `quantum sleeping beauty' scenarios and quantum measurements in very large universes.  Then, in \textsection \ref{pinp} we argue that although in Everettian quantum mechanics there is nothing to be uncertain of before measurement (in idealized cases where the wave function and the dynamics are known), we can still generate rules for rational decision making pre-measurement and theory testing post-measurement.  In \textsection \ref{review} we compare our treatment with the approaches of Zurek, Deutsch, and Wallace.

Before diving in, a choice of terminology.  In this article we'll use `universe' to denote the collection of all branches as described by the universal wave function,  reserving `world' as a way of denoting a particular branch of the wave function, i.e., one of the many worlds of Everettian quantum mechanics.  Using this terminology, `quantum multiverse'$=$`universe.'

\section{Preliminaries: Many-worlds, Self-locating Uncertainty, and Branch-counting}\label{prelim}

\subsection{The Many-worlds Interpretation}\label{introducingEQM}

If the state of the universe is given by a wave function and that wave function's time evolution is at all times in accordance with the Schr\"{o}dinger equation, strange things start to happen.  Many of us have come to accept that tiny particles can be in superpositions of distinct states, e.g., passing through two slits at once.  But, if the wave function always obeys the Schr\"{o}dinger equation then these microscopic superpositions can be amplified, leading to cases where macroscopic measuring devices and even human beings are in superpositions of distinct states.

\begin{description}[font=\normalfont\scshape]
\item[Once] At $t_1$ Alice has prepared particle $a$ in the $x$-spin up eigenstate,
\begin{equation}
\ket{\uparrow_x}_a =  \frac{1}{\sqrt{2}}\Big(\ket{\uparrow_z}_a+\ket{\downarrow_z}_a\Big)\ .
\end{equation}
She then measures the $z$-spin of the particle. At $t_4$, she has just observed the outcome of the experiment.  If the particle were in a $z$-spin eigenstate, Alice and the measuring device would simply record that outcome.  But, since the particle is in a superposition of $z$-spins, Alice and her device evolve into a superposition of different recorded outcomes.  The state of Alice ($A$), her particle ($a$), the measuring device ($D$), and the rest of the universe evolve as below.  Here $R$ indicates the ready state and $\uparrow$/$\downarrow$ denote the possible outcomes.
\begin{align}
\ket{\Psi_{\textsc{Once}}(t_1)}&=\ket{R}_A\ket{R}_{D}\ket{\uparrow_x}_a\ket{E_0}
\nonumber
\\
\ket{\Psi_{\textsc{Once}}(t_4)}&=\frac{1}{\sqrt{2}}\ket{\uparrow}_A\ket{\uparrow}_{D}\ket{\uparrow_z}_a\ket{E_1}+\frac{1}{\sqrt{2}}\ket{\downarrow}_A\ket{\downarrow}_{D}\ket{\downarrow_z}_a\ket{E_2}
\label{once}
\end{align}
\end{description}	

In $\ket{\Psi_{\textsc{Once}}(t_4)}$ both Alice and the detector have entered superpositions of recording up and down as the result of the measurement.  In practice, just one outcome is observed.  One could modify the theory to ensure that only a single outcome actually occurs.  This is the strategy taken by Bohmian mechanics, GRW theory, and other venerable responses to the measurement problem.  However, such a modification may be unnecessary. According to the many-worlds interpretation, $\ket{\Psi_{\textsc{Once}}(t_4)}$ is a state where there are two (or more) copies of Alice, each of which has just observed the readout of a $z$-spin measurement.  There's a version of Alice who saw up and a version who saw down, but no version that saw both up and down.  Seeing a single definite outcome is not precluded by the state evolution given above, it is \emph{guaranteed} since every one of Alice's successors sees a definite outcome.

Even if each copy of Alice has an ordinary experience at $t_4$, one still might be worried about the fact that there exist parts of the universe in which each possible outcome happens.  Particularly, one might hope that after a measurement like this is made and Alice dutifully records up as the result, she can forget about the fact that the particle might have been down for the sake of all her future calculations.  But, if the state of the universe is as in \eqref{once}, then the possibility that the result might have been down is no mere possibility, that part of the wave function is still out there and could potentially interfere with Alice's part.  If the many-worlds interpretation is to be viable, there must be some reason why Alice can ignore the other worlds in which the experiment turned out differently.  Fortunately, there is: decoherence.

As the result of the measurement is recorded in the device and observed by Alice, many traces of this result will appear in the environment.  Due to the large number of traces present in the environment, the states $E_1$ and $E_2$ are (at least approximately) orthogonal and can be expected to stay orthogonal as time progresses.  Because these environment states are and will be orthogonal, the two components of $\ket{\Psi_{\textsc{Once}}(t_4)}$ will evolve in a non-interacting way, each component evolving in accordance with the Schr\"{o}dinger equation as if the other were not present.  They can thus be treated as separate worlds (or collections of worlds) since they are effectively causally isolated and within each of them there are versions of Alice having clear and determinate experiences.\footnote{For more on how Everettian branches arise from decoherence, see \citep[][\textsection 1.3]{schlosshauer2005decoherence, saunders2010} and references therein.  One might object that this sort of story about the emergence of separate worlds cannot be assumed in a derivation of the Born rule.  See appendix \ref{subsys}.}

In quantum mechanics, the state of a subsystem is given by a reduced density matrix, generated by taking a partial trace over the Hilbert space of the environment (\citealp[\textsection 2.4]{schlosshauer}; \citealp[\textsection 2.4]{nielsen2010}). In ordinary cases of quantum measurement, the process of decoherence diagonalizes the reduced density matrix for the macroscopic degrees of freedom describing our observer and measuring apparatus in a very specific basis, the so-called pointer basis (\citeauthor{Zurek:1981xq}, \citeyear{Zurek:1981xq}, \citeyear{Zurek:1982ii}, \citeyear{Zurek:1993ptp}, \citeyear{zurek2003}). The pointer basis is distinguished by being robust with respect to ordinary interactions with the environment.

The state of Alice, the detector, and particle $a$ at $t_1$ can be written as a reduced density matrix by tracing out the environment,
\begin{align}
\widehat{\rho}_{ADa}(t_1)&=\text{Tr}_E\Big(\ket{\Psi_{\textsc{Once}}(t_1)}\bra{\Psi_{\textsc{Once}}(t_1)}\Big)
\nonumber
\\
&=\frac{1}{2}\ket{R}_A\ket{R}_{D}\ket{\uparrow_z}_a\bra{\uparrow_z}_a\bra{R}_{D}\bra{R}_A+\frac{1}{2}\ket{R}_A\ket{R}_{D}\ket{\uparrow_z}_a\bra{\downarrow_z}_a\bra{R}_{D}\bra{R}_A
\nonumber
\\
&\qquad +\frac{1}{2}\ket{R}_A\ket{R}_{D}\ket{\downarrow_z}_a\bra{\uparrow_z}_a\bra{R}_{D}\bra{R}_A+\frac{1}{2}\ket{R}_A\ket{R}_{D}\ket{\downarrow_z}_a\bra{\downarrow_z}_a\bra{R}_{D}\bra{R}_A\ .
\label{equation5}
\end{align}
As the $z$-spin of the particle has yet to be entangled with the measuring device and the environment, the diagonal terms of the reduced density matrix (the middle two terms of the sum in the second line of \eqref{equation5}) are non-zero and potentially very relevant to the evolution of the system---e.g., if the $x$-spin of the particle were to be measured.  After the measurement is made and the environment gets entangled with the particle's spin, the reduced density matrix becomes
\begin{align}
\widehat{\rho}_{ADa}(t_4)&=\frac{1}{2}\bracket{E_1}{E_1}\ket{\uparrow}_A\ket{\uparrow}_{D}\ket{\uparrow_z}_a\bra{\uparrow_z}_a\bra{\uparrow}_{D}\bra{\uparrow}_A+\frac{1}{2}\bracket{E_1}{E_2}\ket{\uparrow}_A\ket{\uparrow}_{D}\ket{\uparrow_z}_a\bra{\downarrow_z}_a\bra{\downarrow}_{D}\bra{\downarrow}_A
\nonumber
\\
&\qquad +\frac{1}{2}\bracket{E_2}{E_1}\ket{\downarrow}_A\ket{\downarrow}_{D}\ket{\downarrow_z}_a\bra{\uparrow_z}_a\bra{\uparrow}_{D}\bra{\uparrow}_A+\frac{1}{2}\bracket{E_2}{E_2}\ket{\downarrow}_A\ket{\downarrow}_{D}\ket{\downarrow_z}_a\bra{\downarrow_z}_a\bra{\downarrow}_{D}\bra{\downarrow}_A
\nonumber
\\
&=\frac{1}{2}\ket{\uparrow}_A\ket{\uparrow}_{D}\ket{\uparrow_z}_a\bra{\uparrow_z}_a\bra{\uparrow}_{D}\bra{\uparrow}_A+\frac{1}{2}\ket{\downarrow}_A\ket{\downarrow}_{D}\ket{\downarrow_z}_a\bra{\downarrow_z}_a\bra{\downarrow}_{D}\bra{\downarrow}_A
\ .
\label{redu}
\end{align}
Here we've assumed for simplicity that $E_1$ and $E_2$ are perfectly orthogonal, $\bracket{E_1}{E_2}=0$, and the diagonal terms vanish; in realistic cases $\bracket{E_1}{E_2}$ is very close to zero.  The diagonal terms have dropped out and the state is now decomposable into two parts---branches---of the reduced density matrix, one in which the result was up and a second in which it was down.  It is common to speak of the universal \emph{wave function} splitting into several branches upon measurement, components that will no longer interact because of decoherence.\footnote{Although we will use the language of `branching' throughout the paper, we do not mean by this to prejudge the issue of whether these branches diverge or overlap (discussed in \citealp{saunders2010b,wilson2012b}).  Our strategy can be implemented on either picture (see footnote \ref{pbuncertainty}).}  Similarly, one can just as easily speak of the \emph{density matrix} for the universe as being decomposable into branches.  \emph{Reduced} density matrices can also evolve into pieces that because of decoherence will no longer interact.  We call these pieces `branches' too, although this is a non-standard use of the term `branch.'  Branches of a reduced density matrix should not be confused with branches of the universal wave function.  A branch of a reduced density matrix may correspond to many branches of the universal wave function, as occurs in the possible elaboration of \textsc{Once} below.

\begin{description}[font=\normalfont\scshape]
\item[Once-or-Twice] Alice's particle ($a$) and Bob's particle ($b$) are both initially prepared in the $x$-spin up eigenstate.  Alice's device measures the $z$-spin of her particle first.  Then, Bob's device, which is connected to Alice's, measures $z$-spin of particle $b$ \emph{only if} particle $a$ was measured to have $z$-spin up.  By $t_1$ the setup is prepared; by $t_2$ Alice's particle has been measured but Bob's has not; by $t_3$ both particles have been measured. Bob has been watching as the results of the experiments are recorded.  Up through $t_3$, Alice has not looked at the measuring devices and is unaware of the results.  By $t_4$, Alice has looked at her device and seen the result of the measurement of particle $a$, although she remains ignorant about the $z$-spin of particle $b$.  The branching structure of this scenario is shown in figure \ref{case1}.  The evolution of the state is shown below.  Here Bob ($B$) and {his }particle {($b$)} have been pulled out of the environment, but otherwise the state is broken down as in \eqref{once} (Bob's device is treated as part of the environment; $\text{X}$ indicates that the measurement was not made).

\begin{align}
\ket{\Psi_{\textsc{OT}}(t_1)}&=\ket{R}_A\ket{R}_{D}\ket{\uparrow_x}_a\ket{R}_B\ket{\uparrow_x}_b\ket{E_{R}}
\nonumber
\\
\ket{\Psi_{\textsc{OT}}(t_2)}&=\frac{1}{\sqrt{2}}\ket{R}_A\ket{\uparrow}_{D}\ket{\uparrow_z}_a\ket{\uparrow}_B\ket{\uparrow_x}_b\ket{E_{\uparrow}}+\frac{1}{\sqrt{2}}\ket{R}_A\ket{\downarrow}_{D}\ket{\downarrow_z}_a\ket{\downarrow}_B\ket{\uparrow_x}_b\ket{E_{\downarrow}}
\nonumber
\\
\ket{\Psi_{\textsc{OT}}(t_3)}&=\frac{1}{2}\ket{R}_A\ket{\uparrow}_{D}\ket{\uparrow_z}_a\ket{\uparrow,\uparrow}_B\ket{\uparrow_z}_b\ket{E_{\uparrow\uparrow}}+\frac{1}{2}\ket{R}_A\ket{\uparrow}_{D}\ket{\uparrow_z}_a\ket{\uparrow,\downarrow}_B\ket{\downarrow_z}_b\ket{E_{\uparrow\downarrow}}
\nonumber
\\
&\qquad +\frac{1}{\sqrt{2}}\ket{R}_A\ket{\downarrow}_{D}\ket{\downarrow_z}_a\ket{\downarrow,\text{X}}_B\ket{\uparrow_x}_b\ket{E_{\downarrow\text{X}}}
\nonumber
\\
\ket{\Psi_{\textsc{OT}}(t_4)}&=\frac{1}{2}\ket{\uparrow}_A\ket{\uparrow}_{D}\ket{\uparrow_z}_a\ket{\uparrow,\uparrow}_B\ket{\uparrow_z}_b\ket{E_{\uparrow\uparrow}'}+\frac{1}{2}\ket{\uparrow}_A\ket{\uparrow}_{D}\ket{\uparrow_z}_a\ket{\uparrow,\downarrow}_B\ket{\downarrow_z}_b\ket{E_{\uparrow\downarrow}'}
\nonumber
\\
&\qquad+\frac{1}{\sqrt{2}}\ket{\downarrow}_A\ket{\downarrow}_{D}\ket{\downarrow_z}_a\ket{\downarrow,\text{X}}_B\ket{\uparrow_x}_b\ket{E_{\downarrow\text{X}}'}
\label{onceortwice}
\end{align}
\end{description}

\begin{figure}[htb]
\center{\includegraphics[width=10cm]{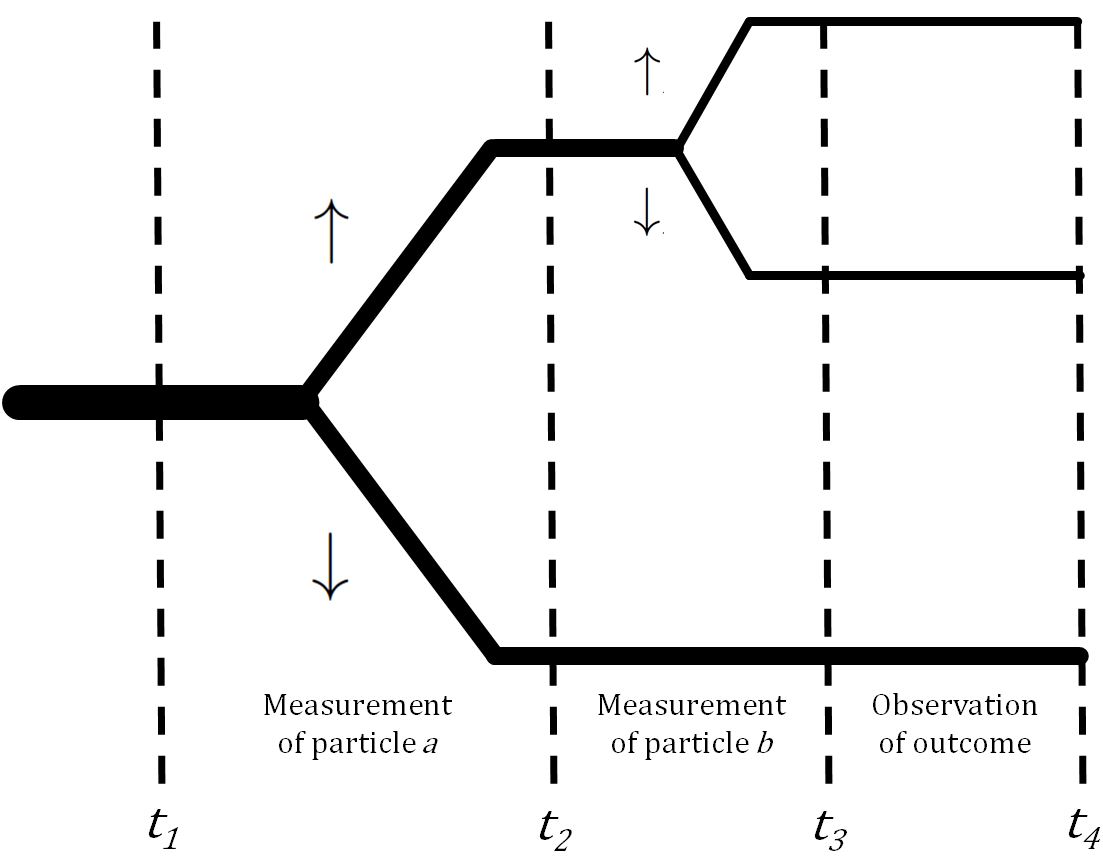}}
\caption{\textsc{Once-Or-Twice}}
\label{case1}
\end{figure}

If we focus on the state of Alice, the detector, and particle $a$ at $t_4$, the reduced density matrix for \textsc{Once-Or-Twice} will be exactly as it was in \textsc{Once} \eqref{redu}.  However, this is clearly a case where there are at least three branches of the universal wave function corresponding to the three different possible outcomes of the series of experiments as recorded by Bob.  In this case, the branch of $\widehat{\rho}_{ADa}$ in which Alice observes up corresponds to at least two branches of the universal wave function, as there are two possible outcomes Bob might have observed.

\subsection{Self-locating Uncertainty and the Everettian Multiverse}\label{sluEM}

On many interpretations of quantum mechanics, quantum measurements yield just one result.  Before a measurement is made, the experimenter is typically uncertain about which outcome will actually occur, even if he knows as much about the universal wave function as could be helpful.  This uncertainty can be quantified using the Born rule.  In the many-worlds interpretation, if we assume that the experimenter knows the relevant information about the wave function, it is unclear what the agent might be uncertain of before a measurement is made.  They know that every outcome will occur and that they will have a successor who sees each possible result.  Before we can discuss the quantitative problem of what numerical probabilities an agent in an Everettian multiverse should assign to different outcomes, we must answer the question: What can one assign probabilities to? Our answer will be that agents performing measurements pass through a period of self-locating uncertainty, in which they can assign probabilities to being one of several identical copies, each on a different branch of the wave function.

Consider Alice in \textsc{Once-or-Twice}. At time $t_2$ the state can be written in either of two equivalent forms,
\begin{align}
\ket{\Psi_{\textsc{OT}}(t_2)}&=\frac{1}{\sqrt{2}}\ket{R}_A\Big(\ket{\uparrow}_{D}\ket{\uparrow_z}_a\ket{\uparrow}_B\ket{\uparrow_x}_b\ket{E_{\uparrow}}+\ket{\downarrow}_{D}\ket{\downarrow_z}_a\ket{\downarrow}_B\ket{\uparrow_x}_b\ket{E_{\downarrow}}\Big)
\label{oot1}
\\
&=\frac{1}{\sqrt{2}}\ket{R}_A\ket{\uparrow}_{D}\ket{\uparrow_z}_a\ket{\uparrow}_B\ket{\uparrow_x}_b\ket{E_{\uparrow}}+\frac{1}{\sqrt{2}}\ket{R}_A\ket{\downarrow}_{D}\ket{\downarrow_z}_a\ket{\downarrow}_B\ket{\uparrow_x}_b\ket{E_{\downarrow}}.
\label{oot2}
\end{align}
In the first way of writing the state, it appears as if there is only one copy of Alice (represented by $\ket{R}_A$), while the rest of the state has branched in two. But in the second expression, it appears as if there are two branches with two identical copies of Alice. \eqref{oot1} and \eqref{oot2} are manifestly mathematically identical, but suggest different physical pictures, corresponding to two plausible attitudes about wave function branching. One attitude would be that Alice herself doesn't branch until her own reduced density matrix splits into multiple distinct branches (in the pointer basis) sometime between $t_3$ and $t_4$; in that case \eqref{oot1} gives the right physical picture, and there is only one copy of Alice even after the first measurement has occurred. The other attitude is that branching happens throughout the whole wave function whenever it happens anywhere.  When the universal wave function splits into multiple distinct and effectively non-interacting parts, the entire world splits---along with every object and agent in it. Then \eqref{oot2} is the more perspicuous way of writing the state.  There are two identical post-measurement copies of Alice, one on each branch.\footnote{\citet[\textsection 8.6]{wallace2012} advocates a version of the first picture and \citeauthor{vaidman2014} (\citeyear{vaidman2014}, \citeyear{vaidmanSEP}, \citeyear{vaidman2015}) the second.}

We will work under the assumption that the second attitude (branching happens throughout the wave function whenever it happens anywhere) is the right way to think about the process.\footnote{In our derivation of the Born rule, \textsection \ref{prf}, this second attitude plays an important role in allowing us to examine cases like \eqref{state15050} where the agent is in a physically identical state on each branch.} The branching structure of the wave function in Everettian quantum mechanics is not uniquely defined by the theory, which is just a state space and a dynamical law; like coarse-graining in classical statistical mechanics, there is a subjective element in how it is constructed.  We must decide what kind (and what degree) of separation is required for two putative branches to count as distinct (this involves choosing an appropriate system/environment split, or, in a consistent-histories approach, choosing which coarse-grained observables to use). However, treating branching on a subsystem-by-subsystem basis---as advised by the first attitude---is needlessly arbitrary; it depends on how the universe is carved into subsystems. Some very small piece of a post-measurement detector might remain in the same state on every branch, but it seems wrong to think that whether the single piece has branched into many depends on whether it is considered as part of the detector or as a separate subsystem in its own right. For Alice's purposes, it makes sense to distinguish between the two copies of herself in \eqref{oot2}; even though they are currently unentangled with the rest of the state, in the future they will evolve differently (and independently of one another) as Alice becomes aware of the measurement outcome. (In particular, she is \emph{allowed} to make that distinction, and doing so enables her to assign probabilities to measurement outcomes; probabilities which we will show must obey the Born rule.)

The non-local nature of the globally-branching view might cause some discomfort. It implies that observers here on Earth could be (and almost surely are) branching all the time, without noticing it, due to quantum evolution of systems in the Andromeda Galaxy and elsewhere throughout the universe.\footnote{However, the reduced density matrix describing the state of the observer on Earth is unaffected by the events occurring in the Andromeda galaxy.  In a relativistic context, one natural option would be to say that although the number of copies of an observer is frame-relative, the probabilities assigned to measurement outcomes are not (as the reduced density matrix is all that is relevant to the calculation of probabilities, see \textsection \ref{sec21}).} We take this to be one among many psychologically unintuitive but empirically benign consequences of Everettian quantum mechanics.

There are two copies of Alice at $t_2$ in \textsc{Once-or-Twice}. Each copy can reasonably wonder which one she is. Thus even if she (incredibly) knows the universal wave function exactly, Alice still has something to be uncertain of.  She isn't uncertain about the way the universe is; by supposition she knows the wave function and this gives a complete specification of the state of the universe.  Alice is uncertain about where she is in the quantum multiverse (as has been emphasized by \citeauthor{vaidman1998} \citeyear{vaidman1998}, \citeyear{vaidman2011}, \citeyear{vaidman2014}, \citeyear{vaidmanSEP}).  She doesn't know if she's in the branch of the wave function in which the detector displays up or the one in which it shows down.  We say that Alice has \emph{self-locating uncertainty} \citep[see][]{lewis1979, bostrom2002}.  We call this period in which self-locating uncertainty is present, after the measurement has been made and branching has occurred via decoherence but before the experimenter has registered the result, the `post-measurement pre-observation' period.

During the post-measurement pre-observation there are multiple copies of Alice seeing different results.  Are any or all of them the same person as Alice before the measurement?  This is a tricky metaphysical question upon which we will not speculate.  We will indiscriminately refer to a post-branching version of Alice as a `copy,' or a `successor,' or simply `Alice.'  In appealing to self-locating uncertainty during the post-measurement pre-observation period, all that is needed is that each copy of Alice can coherently wonder what sort of branch they inhabit.

At this point it may be objected that \textsc{Once-or-Twice} is highly atypical as Alice is for a time unaware of the measurement result.  Actually, self-locating uncertainty is generic in quantum measurement. In Everettian quantum mechanics the wave function branches when the system becomes sufficiently entangled with the environment to produce decoherence.  The normal case is one in which the quantum system interacts with an experimental apparatus (cloud chamber, Geiger counter, electron microscope, or what have you) and then the observer sees what the apparatus has recorded. For any realistic room-temperature experimental apparatus, the decoherence time is extremely short: less than $10^{-20}$ seconds. Even if a human observer looks at the quantum system directly, the state of the observer's eyeballs will decohere in a comparable time. In contrast, the time it takes a brain to process a thought is measured in tens of milliseconds. No matter what we do, real observers will find themselves in a situation of self-locating uncertainty (after decoherence, before the measurement outcome has been registered).\footnote{Although we think that self-locating uncertainty is typical, it may not be necessary for our account that it is always present.  See the end of \textsection \ref{inflink}.}  The observer may not have enough time to think and reorganize their credences before learning the outcome \citep[\textsection 4.2]{wallace2006}, but, in trying to approximate ideal rationality, the agent can attempt to reconstruct the probabilities they ought to have assigned during the post-measurement pre-observation period.  Despite often being short-lived, these probabilities are relevant for deciding what to believe after seeing the measurement outcome, \textsection \ref{inflink}, and deciding how to act in anticipation of the measurement, \textsection \ref{betandbranch}.

Self-locating uncertainty has been discussed in great depth by formal epistemologists.  The general question of how to extend Bayesian updating to cases involving self-locating uncertainty is very difficult and a matter of contemporary philosophical debate, but many different proposals\footnote{In particular, we're thinking of the proposals of \citep{bostrom2002,meacham2008,ManleyF} as formulated in \citep{ManleyF}.}\label{listofproposals} agree on one minimal constraint on rational credences (a.k.a. subjective probabilities or degrees of belief):
\begin{description}[font=\normalfont\bfseries]
\item[Indifference] Suppose that in a given possible universe $U$, agent $S$ has a finite number $N_U$ of copies in an internally qualitatively identical state to $S$.  Further, suppose that the hypothesis $H$ is true for $N_{UH}$ of these copies.  Then, $S$ should assign the following conditional credence to $H$ being true given that $S$ is in $U$:
\begin{equation}
P(H|U)=\frac{N_{UH}}{N_U}
\label{}
\end{equation}
\end{description}
\emph{Indifference} was originally put forward in \citep{elga2004}; the version here is from \citep{ManleyF} (paraphrased).  Two agents are in the same \emph{internal qualitative state} if they have identical current evidence: the patterns of colors in their visual fields are identical, they recall the same apparent memories, they both feel equally hungry, etc.  This principle should not be confused with a Laplacean `principle of indifference.'

To see \emph{Indifference} in action, consider the following case from \citet{elga2004}:
\begin{description}[font=\normalfont\scshape]
\item[Duplicating Dr. Evil]  Dr. Evil is plotting the destruction of Earth from his lunar battle station when he receives an unwelcome message.  Back on earth some pesky philosophers have duplicated the entirety of his battle station, perfectly replicating every piece of furniture, every weapon, and every piece of food, even replicating the stale moon air and somehow the weaker gravitational field.  They went so far that at $t$ they created a duplicate of Dr. Evil himself, Dup.  Dup's immediate experiences are internally identical to Dr. Evil's.  The two men pace, think, and scratch themselves in perfect synchronicity.
\end{description}
Upon learning of his duplication, what credence should Dr. Evil assign to being in his lunar base and not the terrestrial fake?  According to \textit{Indifference}, Dr. Evil should assign a credence of one half (and Dup should do the same).  Dr. Evil knows that in any universe $U$ he might inhabit there are two copies of himself ($N_U=2$) and that one is on the moon ($N_{UH}=1$ for $H=\text{`I am on the moon.'}$).

\subsection{Indifference and the Quantitative Probability Problem}\label{quant}

In the previous section we showed that in the post-measurement pre-observation period it is possible to assign probabilities to the different possible outcomes of a completed experiment.  But, what probabilities ought one to assign?  If you take \emph{Indifference} to be a constraint on rational probability assignments and count agents in the most naive way, there appears to be a simple rule for assigning probabilities: branch-counting.\footnote{\citet{lewis2009} considers a proposal along these lines.}  The probability that $H$ is true is given by the fraction of branches in which $H$ holds.  Since there is a single copy of the experimenter on each branch, the fraction of branches is the same as the fraction of agents and branch-counting is a direct consequence of \emph{Indifference}.

There is a well-known problem for this kind of approach.\footnote{For more in-depth discussion, see \citeauthor{wallace2007} (\citeyear{wallace2007}, \textsection 9, \citeyear{wallace2012}, \textsection 3.11).}  The number of branches in which a certain outcome occurs might not be well-defined.  Branches are structures in the wave function which emerge via decoherence, and although this process will guarantee that on each branch a definite outcome is observed, it will generally not yield a good answer to the question of how many branches feature a certain outcome or even how many branches there are in total.  In contrast, the total mod-squared amplitude of branches in which a certain outcome occurs is well-defined (although not to arbitrary precision).  The difficulty of branch-counting provides good reason to think \emph{Indifference} will often be unhelpful in Everettian scenarios.  However, \emph{Indifference} does give clear recommendations in idealized scenarios, so for the time being we'll restrict our attention to cases where the number of branches is well-defined.

Assume that \textsc{Once-or-Twice} really is a case in which one branch becomes three, as depicted in figure \ref{case1}.  At $t_3$, what probability should Alice assign to particle $a$ having been measured to be $z$-spin down?  By the methods of textbook quantum mechanics, we should use the Born rule to calculate the probability.  That is, we should square the coefficient of the third branch of the wave function in \eqref{onceortwice},
\begin{equation}
P_{t_3}(\text{down})=\left(\frac{1}{\sqrt{2}}\right)^2=\frac{1}{2}\ .
\end{equation}
To apply \emph{Indifference} we simply need to determine the fraction of branches in which the result of the first measurement was down,
\begin{equation}
P_{t_3}(\text{down})=\frac{\text{number of down branches}}{\text{number of branches}}=\frac{1}{3}\ ,
\label{indifferencet3}
\end{equation}
the coefficients of $1/2$, $1/2$, and $1/\sqrt{2}$ of the branches in $\ket{\Psi_{\textsc{OT}}(t_3)}$ are irrelevant to the calculation.  Supposing \emph{Indifference} is true and people are to be counted in this straightforward way, textbook quantum mechanics and Everettian quantum mechanics make different predictions about the probability of seeing down.  Thus, we can empirically test the many-worlds interpretation in any case where the fraction of branches in which some hypothesis $H$ is true is unequal to the combined weights of branches in which $H$ is true (the generic case).

We now have the strongest kind of argument against Everettian quantum mechanics: it is empirically inadequate.  The theory fails to reproduce the empirical predictions of textbook quantum mechanics and thus has been conclusively falsified by the data typically taken as evidence for quantum mechanics.\footnote{Admittedly, the force of this argument is weakened by the fact that we are working under an idealization.  Branch number may not be well-defined in realistic cases and then it becomes unclear what probabilities the proponent of \emph{Indifference} thinks one should assign.}  However, this argument relies on the assumption that \emph{Indifference} is a universally valid principle applicable to Everettian multiverses as well as cases of classical duplication like \textsc{Duplicating Dr. Evil}.  As we will see, it is not.

\subsection{Against Branch-counting}\label{response}

When \emph{Indifference} is applied to \textsc{Once-or-Twice}, something odd happens.  At $t_2$ the principle advises each copy of Alice to assign a probability of one half to particle $a$ being down as half of the copies of Alice are in a branch where the particle was measured to be down.  Then, at $t_3$ Alice is supposed to assign down a probability of one third (as in \eqref{indifferencet3}).  According to the Born rule, the recommended probability assignment is one half at both $t_2$ and $t_3$.  If \emph{Indifference} is right, there's a strange switch in the probabilities between $t_2$ and $t_3$.  Is there any reason to think this undermines the branch-counting strategy advocated by \emph{Indifference}?  David Wallace has argued that such a switch violates a constraint he calls `diachronic consistency.' In appendix \ref{notaproblem}, we argue that this is not the right diagnosis of the problem with the switch in credences.  This kind of `inconsistency' is a common result of \emph{Indifference} and not something which should be taken to refute the principle.  Still, we agree that there's something wrong with the probability switch.

Between $t_2$ and $t_3$ what happens?  Particle $b$ is measured and Bob takes note of the result.  Nothing happens to Alice, particle $a$, or Alice's device.  If nothing about Alice or her detector changes, why should her degree of belief that she bears a certain relation to the detector change?  At $t_2$, the state of Alice and her detector is given by,
\begin{equation}
\widehat{\rho}_{AD}(t_2)=\frac{1}{2}\ket{R}_A\ket{\uparrow}_{D}\bra{\uparrow}_{D}\bra{R}_A+\frac{1}{2}\ket{R}_A\ket{\downarrow}_{D}\bra{\downarrow}_{D}\bra{R}_A\ .
\label{reddensAD}
\end{equation}
At this time Alice is trying to locate herself in one of the two branches of this Alice+Detector system and, if she applies \emph{Indifference}, decides that she should assign a probability of one half to each branch.  At $t_3$, the state of the Alice+Detector system is unchanged, it is still exactly as in \eqref{reddensAD}.  Now she decides that the probability of being in the second branch is one third.  Why should her probability for being in different subsystems (branches of $\widehat{\rho}_{AD}$) of the Alice+Detector system change when nothing about that system changes and she knows that she is somewhere in that system?  It shouldn't.  In the next section we will use this insight to motivate a replacement for \emph{Indifference}.

It is tempting to think that the number of copies of Alice cannot change without her physical state changing---this is the way things work in classical physics.  But, in Everettian quantum mechanics, changes that purely affect her environment can change the number of copies of Alice in existence.  For example, the change of state from $t_2$ to $t_3$ in \textsc{Once-or-Twice}, \eqref{onceortwice}.  Two intuitive constraints come into conflict: \emph{Indifference}, and the belief that Alice's probabilities should be unaffected by changes in the state of her environment.  We recommend rejecting the former in favor of the latter.

\section{The Epistemic Irrelevance of the Environment}\label{newf}
\subsection{The Epistemic Separability Principle}\label{sec21}

In \textsection \ref{response} we argued that branch-counting is unreasonable because it requires Alice to change her credences about the result of the measurement of particle $a$ when things change elsewhere in the universe. In particular, when particle $b$ is measured.  In the \textsc{Duplicating Dr. Evil} case, whatever one thinks of \emph{Indifference}, it seems clear that certain facts are irrelevant. The probability $P(\text{I'm Dr. Evil}|U)$ should not depend on what's happening deep inside the Earth's core or what's happening on the distant planet Neptune or any other remote occurrences.  Unless, of course, the actions of the Earth's core cause earthquakes which the terrestrial Dup feels but the lunar Dr. Evil does not.  Or, if what's happening on Neptune includes \emph{another} copy of the laboratory with another duplicate Dr. Evil in it.  If there's a duplicate on Neptune, Dr.Evil can no longer be sure that Neptune is in fact a distant planet and not the one under his feet (and thus cannot treat it as irrelevant to his probability assignments).  As long as the copies of Dr. Evil are unaffected and no new copies are created elsewhere, $P(\text{I'm Dr. Evil}|U)$ should be unaffected by changes to the environment.  This thought, which was essential to our argument in \textsection \ref{response}, can be stated in imprecise slogan form as:
\begin{description}
\item[ESP-gist]  The credence one should assign to being any one of several observers having identical experiences is independent of the state of the environment.
\end{description}
In more precise but still theory-independent language, this epistemic principle becomes:
\begin{description}
\item[ESP]  Suppose that universe $U$ contains within it a set of subsystems $\mathcal{S}$ such that every agent in an internally qualitatively identical state to agent $A$ is located in some subsystem which is an element of $\mathcal{S}$.  The probability that $A$ ought to assign to being located in a particular subsystem $X\in\mathcal{S}$ given that they are in $U$ is identical in any possible universe which also contains subsystems $\mathcal{S}$ in the same exact states (and does not contain any copies of the agent in an internally qualitatively identical state that are not located in $\mathcal{S}$).
\begin{equation}
P(X|U)=P(X|\mathcal{S})
\label{esp}
\end{equation}
\end{description}
ESP stands for `Epistemic Separability Principle,' as the principle allows one to separate the relevant parts of the universe from the rest; that is, to separate the set of subsystems where the agent might, for all they know, be located from everything else. $X$, $U$, and $\mathcal{S}$ are not propositions; \eqref{esp} is shorthand that must be clarified.  $P(X|U)$ is the probability that $A$ assigns to being \emph{in} $X$ given that they are \emph{in} $U$.  In $P(X|\mathcal{S})$, `$\mathcal{S}$' is shorthand for: there exist subsystems $\mathcal{S}$ and there are no internally qualitatively identical copies of $A$ outside of these subsystems.  To be precise, let's say that one is `located in' subsystem $X$ just in case one is a proper part of the subsystem (no limb or brain cell is omitted).  There may be more than one copy of $A$ in a given subsystem.  In general, There will be many ways to carve out a set of subsystems from the universe.  \emph{ESP} applies to any such carving.  The subsystems in $\mathcal{S}$ together need not cover the entire world, there may well be parts of $U$ that are not in any of these subsystems, however these omitted parts cannot contain copies of $A$.  The principle is restricted to cases where $\mathcal{S}$ has a finite number of members.  The subsystems need not be located at the same time.  One may, for example, be unsure if they are the person waking up in their bed on Monday or Tuesday (see \textsection \ref{mixed}).  

We need to be somewhat careful about what exactly a `subsystem' is supposed to be, although we will not attempt to give a rigorous characterization applicable to any conceivable theory. The essential idea is that a subsystem is a part of the larger system that can be considered as a physical system in its own right. Slightly more formally, we imagine that the overall state of a system can be decomposed into the states of various subsystems, so that two constraints are satisfied: (1) the state of each subsystem, perhaps with some additional information about how the subsystems are connected, can be used to uniquely reconstruct the original state; and (2) the information contained within each subsystem's state is enough to specify its immediate dynamical evolution, as long as the other subsystems are not influencing it.\footnote{Work would need to be done to formulate these constraints precisely, here they serve as rough guides.} So, in classical particle physics, a subsystem might be a collection of particles at a time and its state specified by giving the masses, positions, and velocities of the particles in the collection.  By contrast, giving only the positions (or only the velocities) of a collection of particles would not specify a subsystem, since that wouldn't be sufficient to determine the collection's evolution. In quantum mechanics, systems can be divided into subsystems in two fundamentally different ways---quite unlike classical physics.  The state of a system can be written as a density matrix $\widehat{\rho}_{sys}$ corresponding to some factor of the Hilbert space $\mathscr{H}_{sys}$ (a \emph{reduced} density matrix unless the system is the whole universe).  As $\mathscr{H}_{sys}$ can itself be decomposed into factors, $\mathscr{H}_{sys}=\mathscr{H}_1\otimes\mathscr{H}_2$, one can treat the reduced density matrices $\widehat{\rho}_1$ and $\widehat{\rho}_2$ as subsystems of $\widehat{\rho}_{sys}$.  Alternatively, if---as in \eqref{reddensAD}---branching has occurred, the different branches of $\widehat{\rho}_{sys}$ can be regarded as separate subsystems.\footnote{In specifying the state of each branch of a reduced density matrix like \eqref{reddensAD} one must retain the numerical prefactors before each term as they would be necessary to reconstruct the state of the system as a whole.}

Examination of the argument given by \citet{elga2004} for \emph{Indifference} reveals that something like \emph{ESP} is taken for granted. In his \textsc{Toss\&Duplication} thought experiment, Elga assumes that the outcome of an additional coin toss should not affect the credence we assign to being either the original or a duplicated person with identical experiences; the justification for such an assumption would have to be something like \emph{ESP}. \emph{ESP} is compatible with \emph{Indifference} in standard cases of classical self-locating uncertainty like \textsc{Duplicating Dr. Evil}.  Requiring that all one care about in assigning credences between Dr. Evil and Dup is what's happening in the lunar laboratory, $X$, and the terrestrial replica, $Y$, (together, $\mathcal{S}=\{X,Y\}$) looks like it allows any assignment of credences to the two copies at all, provided one is consistent across universes that vary only in the state of the world outside the two laboratories.  Actually, in \textsection \ref{ESPIndifference} we'll see that although \emph{ESP} is compatible with \emph{Indifference}, the rule is not as permissive as it might initially seem.  At least, not if we strengthen it as in \textsection \ref{ESPIndifference}.  In classical cases, the strengthened principle requires one to consider each copy of oneself to be equiprobable---in agreement with \emph{Indifference}.

Now that \emph{ESP} is on the table, let's apply it to cases of measurement in Everettian quantum mechanics.  Consider \textsc{Once-or-Twice}.  At $t_2$, after particle $a$ has been measured, Alice has branched into multiple copies.  Alice knows that she is somewhere in the Alice+Detector system which is characterized by the reduced density matrix in \eqref{reddensAD}.  The density matrix $\widehat{\rho}_{AD}$ itself can be divided into branches $X$ \& $Y$ corresponding to the particle spin being measured as either up or down.  Any universal wave function which agrees on the density matrix $\widehat{\rho}_{AD}$ agrees on the state of $X$ and $Y$.  So, the probability of being in $X$, which is the probability of an up result, must be the same in any universe with the same reduced density matrix $\widehat{\rho}_{AD}$.

More generally, suppose that an experimenter $A$ has just measured observable $\widehat{O}$ of system $S$ and the measuring device has recorded some eigenvalue $O_i$ on each branch of the wave function.\footnote{\emph{ESP} is not \emph{only} applicable to Everettian quantum mechanics in cases of measurement.  It can be applied whenever one is trying to locate oneself within a collection of decohered branches of the wave function.  For example, this decoherence might have been caused by chaotic processes instead of quantum measurement \citep[ch. 3]{wallace2012}.}  As discussed in \textsection \ref{introducingEQM}, the reduced density matrix, $\widehat{\rho}_{AD}$, will be diagonalized in the pointer basis for the Agent+Detector subspace, $\mathscr{H}_A\otimes \mathscr{H}_D$, by the decoherence process.  Each pointer state (with nonzero amplitude) defines a branch of the reduced density matrix, $\widehat{\rho}_{AD}$, on which the detector $D$ registered a particular outcome $O_i$.\footnote{After a measurement, the branching structure of $\widehat{\rho}_{AD}$---the set of pointer states---can be derived either by examining the interaction with the environment or by seeing in which basis the matrix is diagonal.  One need not know anything about the state of the environment; the reduced density matrix alone is sufficient.}  In assigning credences to the different outcomes, the agent is assigning probabilities to being located in these different branches.  Specifying the state of $\widehat{\rho}_{AD}$ determines the state of all its branches and thus of all the subsystems in which the agent might find themselves ($\mathcal{S}$).

Thus in ordinary cases of quantum measurement, we can formulate a less general version of \emph{ESP} which will be sufficient for our derivation of the Born rule.
\begin{description}
\item[ESP-QM]  Suppose that an experiment has just measured observable $\widehat{O}$ of system $S$ and registered some eigenvalue $O_i$ on each branch of the wave function.  The probability that agent $A$ ought to assign to the detector $D$ having registered $O_i$ when the universal wave function is $\Psi$, $P(O_i|\Psi)$, only depends on the reduced density matrix of $A$ and $D$, $\widehat{\rho}_{AD}$:
\begin{equation}
P(O_i|\Psi)=P(O_i|\widehat{\rho}_{AD})
\end{equation}
\end{description}
This principle tells us that when observers assign probabilities to recorded outcomes of measurements that have already occurred, these probabilities should only depend on the Agent+Detector state, $\widehat{\rho}_{AD}$ (not on other features of the universal wave function).  By applying this principle to quantum cases instead of \emph{Indifference}, we are now able to shake the unrealistic assumption that the number of branches in which a certain outcome occurs is well-defined.

In formulating \emph{ESP-QM}, we've relied on the fact that the state of a quantum subsystem is specified by a reduced density matrix.  Although this is the standard way of representing subsystems in quantum mechanics, one might worry that its use here requires further justification.  We discuss this concern in appendix \ref{subsys}.

\subsection{Deriving the Born Rule}\label{prf}

In this section we will derive the Born rule probabilities as the rational assignment of credences post-measurement pre-observation.  We will first derive the rule in a case with two branches that have equal amplitudes, then use similar techniques to treat a case with two branches of unequal amplitude.  It is straightforward to extend these methods to more general cases (see appendix \ref{generalization}). Mathematically, our argument is most similar to that of \citet{zurek2005} and not far from those of the decision theoretic approach (\citealp{deutsch1999}; \citeauthor{wallace2003b}, \citeyear{wallace2003b}, \citeyear{wallace2010b}, \citeyear{wallace2012}).  The interest of this proof is not its mathematical ingenuity but the facts that (a) it applies to cases where uncertainty is undeniably present and (b) it is based on a single well-motivated principle of rationality, \emph{ESP}.

\textbf{Proof of the Born rule for $\frac{1}{2}$/$\frac{1}{2}$ case}: Alice measures the $z$-spin of a single particle in the $x$-spin up state.  One display ($D1$) will show the result of the experiment.  If the spin is up, a second display ($D2$) will show $\heartsuit$.  If it is down, a $\diamondsuit$ will appear on the second display.  Alice is not immediately affected by the result; in particular, she is for a time unaware of the experiment's outcome.\footnote{That is, we assume that Alice is in the same physical state on both branches of the wave function ($\ket{R}_A$ in \eqref{state15050}).  As time passes and the result of the measurement has effects on various parts of the universe, Alice's state will come to differ on the two branches (e.g., if she observes the result).  But, immediately after measurement her state will be unaffected.} The wave function of Alice, the detectors, the particle, and the environment (the rest of the universe) evolves from
\begin{equation}
\ket{\Psi_0}=\ket{R_0}_A\ket{R}_{D1}\ket{R}_{D2}\ket{\uparrow_x}\ket{E_R}
\end{equation}
to
\begin{equation}
\ket{\Psi_1}=\frac{1}{\sqrt{2}}\ket{R}_A\ket{\uparrow}_{D1}\ket{\heartsuit}_{D2}\ket{\uparrow_z}\ket{E_{\uparrow \heartsuit}}+\frac{1}{\sqrt{2}}\ket{R}_A\ket{\downarrow}_{D1}\ket{\diamondsuit}_{D2}\ket{\downarrow_z}\ket{E_{\downarrow\diamondsuit}}\ .
\label{state15050}
\end{equation}
To use \emph{ESP-QM} to demonstrate that $P\left(\uparrow\middle|\Psi_1\right)=P\left(\downarrow\middle|\Psi_1\right)=1/2$, we will need to also consider an \textit{alternate scenario} where the computer (part of the environment) is programmed differently so that $\heartsuit$ displays if down is measured and $\diamondsuit$ displays if up.  Then the post-measurement pre-observation wave function would be
\begin{equation}
\ket{\Psi_2}=\frac{1}{\sqrt{2}}\ket{R}_A\ket{\uparrow}_{D1}\ket{\diamondsuit}_{D2}\ket{\uparrow_z}\ket{E_{\uparrow \diamondsuit}}+\frac{1}{\sqrt{2}}\ket{R}_A\ket{\downarrow}_{D1}\ket{\heartsuit}_{D2}\ket{\downarrow_z}\ket{E_{\downarrow \heartsuit}}\ .
\label{state25050}
\end{equation}
\textbf{Step 1}:  Focus first on Alice and $D1$.  The Alice+Detector $1$ reduced density matrices for $\Psi_1$ and $\Psi_2$ are the same,\footnote{Here we assume for simplicity that the environments are perfectly orthogonal.  What is crucial is that we choose $\Psi_2$ such that $\widehat{\rho}_{AD1}\left(\Psi_1\right)=\widehat{\rho}_{AD1}\left(\Psi_2\right)$ and $\widehat{\rho}_{AD2}\left(\Psi_1\right)=\widehat{\rho}_{AD2}\left(\Psi_2\right)$, both of which can be easily satisfied even if the environment states are not perfectly orthogonal.}
\begin{equation}
\widehat{\rho}_{AD1}\left(\Psi_1\right)=\widehat{\rho}_{AD1}\left(\Psi_2\right)=\frac{1}{2}\ket{R}_A\ket{\uparrow}_{D1}\bra{R}_A\bra{\uparrow}_{D1}+\frac{1}{2}\ket{R}_A\ket{\downarrow}_{D1}\bra{R}_A\bra{\downarrow}_{D1}\ .
\label{d1}
\end{equation}
\emph{ESP-QM} requires that the probabilities Alice assigns to the possible spin results be the same in these two universes as they have the same Observer+Detector reduced density matrix,
\begin{equation}
P\left(\downarrow\middle|\Psi_1\right)=P\left(\downarrow\middle|\Psi_2\right)\ .
\label{s1}
\end{equation}
\textbf{Step 2}: If we ask what probability Alice should assign to the display being $\heartsuit$, we need to consider the reduced density matrix generated by tracing over $D1$, the spin of the particle, and the environment.  $\Psi_1$ and $\Psi_2$ agree on $\widehat{\rho}_{AD2}$.  By \emph{ESP-QM}, the probabilities assigned to $\heartsuit$ must be equal,
\begin{equation}
P\left(\heartsuit\middle|\Psi_1\right)=P\left(\heartsuit\middle|\Psi_2\right)\ .
\label{s2}
\end{equation}
\textbf{Step 3}: Next, note that the $\heartsuit$-branches \textit{just are} the $\uparrow$-branches in $\Psi_1$ and the $\heartsuit$-branches \textit{just are} the $\downarrow$-branches in $\Psi_2$.  Thus Alice is in the $\heartsuit$-branch of $\widehat{\rho}_{AD2}(\Psi_1)$ if and only if she is in the $\uparrow$-branch of $\widehat{\rho}_{AD1}(\Psi_1)$.  Similarly, she is in the $\heartsuit$-branch of $\widehat{\rho}_{AD2}(\Psi_2)$ if and only if she is in the $\downarrow$-branch of $\widehat{\rho}_{AD1}(\Psi_2)$.   Therefore, Alice must assign
\begin{align}
P\left(\uparrow\middle|\Psi_1\right)&=P\left(\heartsuit\middle|\Psi_1\right)
\nonumber
\\
P\left(\downarrow\middle|\Psi_2\right)&=P\left(\heartsuit\middle|\Psi_2\right)\ .
\label{s3}
\end{align}
\textbf{Step 4}:  Putting together the results in \eqref{s1} - \eqref{s3}, we see that in $\Psi_1$ the probability of being on a $\uparrow$/$\heartsuit$-branch must be the same as that for being on a $\downarrow$/$\diamondsuit$-branch: $P\left(\uparrow\middle|\Psi_1\right)=P\left(\downarrow\middle|\Psi_1\right)$.  So, the unique rational degrees of belief in the first scenario consider each branch to be equiprobable.  Since these are the only two alternatives, the probability of each outcome is one half.  In fact, using \emph{ESP-QM}, we have shown that for \textit{any} case in which the reduced density matrix is as in \eqref{d1}, the two spin states are equiprobable; there doesn't have to be a second display present.  The result applies to a state in the general form:
\begin{equation}
\frac{1}{\sqrt{2}}\ket{R}_A\ket{\uparrow}_{D1}\ket{\uparrow_z}\ket{E_{\uparrow}}+\frac{1}{\sqrt{2}}\ket{R}_A\ket{\downarrow}_{D1}\ket{\downarrow_z}\ket{E_\downarrow}
\label{general5050}
\end{equation}

\textbf{Proof of the Born rule for $\frac{1}{3}$/$\frac{2}{3}$ case}:  Suppose Alice measures a particle in the state
\begin{equation}
\sqrt{\frac{2}{3}}|\uparrow_z\rangle+\sqrt{\frac{1}{3}}|\downarrow_z\rangle\ ,
\end{equation}
in which case upon measurement $\widehat{\rho}_{AD1}$ would be
\begin{equation}
\widehat{\rho}_{AD1}\left(\Psi_2\right)=\frac{2}{3}\ket{R}_A\ket{\uparrow}_{D1}\bra{R}_A\bra{\uparrow}_{D1}+\frac{1}{3}\ket{R}_A\ket{\downarrow}_{D1}\bra{R}_A\bra{\downarrow}_{D1}\ .
\label{23reduced}
\end{equation}
To determine the probabilities in this scenario, we will consider two different ways of having three displays linked to the measurement outcomes in the post-measurement pre-observation state,
\begin{align} \ket{\Psi_\alpha}&=\sqrt{\frac{1}{3}}\Big(\ket{R}_A\ket{\uparrow}_{D1}\ket{\diamondsuit}_{D2}\ket{\clubsuit}_{D3}\ket{\uparrow_z}\ket{E_{\alpha 1}}+\ket{R}_A\ket{\uparrow}_{D1}\ket{\heartsuit}_{D2}\ket{\spadesuit}_{D3}\ket{\uparrow_z}\ket{E_{\alpha 2}}
\nonumber
\\
&\qquad
+\ket{R}_A\ket{\downarrow}_{D1}\ket{\heartsuit}_{D2}\ket{\clubsuit}_{D3}\ket{\downarrow_z}\ket{E_{\alpha 3}}\Big)
\nonumber
\\
\ket{\Psi_\beta}&=\sqrt{\frac{1}{3}}\Big(\ket{R}_A\ket{\uparrow}_{D1}\ket{\heartsuit}_{D2}\ket{\clubsuit}_{D3}\ket{\uparrow_z}\ket{E_{\beta 1}}+\ket{R}_A\ket{\uparrow}_{D1}\ket{\heartsuit}_{D2}\ket{\clubsuit}_{D3}\ket{\uparrow_z}\ket{E_{\beta 2}}
\nonumber
\\
&\qquad+\ket{R}_A\ket{\downarrow}_{D1}\ket{\diamondsuit}_{D2}\ket{\spadesuit}_{D3}\ket{\downarrow_z}\ket{E_{\beta 3}}\Big)
\label{alphaandbetastates}
\end{align}
\textbf{Step 1}:  Using \emph{ESP-QM} to ignore $D2$, $D3$, and the environment, we can focus on the first display and compare the probabilities for $\downarrow$,
\begin{equation}
P\left(\downarrow\middle|\Psi_\alpha\right)=P\left(\downarrow\middle|\Psi_\beta\right)\ .
\label{ss1}
\end{equation}
\textbf{Step 2}:  Focusing on the second display gives
\begin{equation}
P\left(\diamondsuit\middle|\Psi_\alpha\right)=P\left(\diamondsuit\middle|\Psi_\beta\right)\ . 
\label{ss15}
\end{equation}
Since the $\diamondsuit$-branches of $\Psi_\beta$ are the $\downarrow$-branches, we have
\begin{equation}
P\left(\diamondsuit\middle|\Psi_\beta\right)=P\left(\downarrow\middle|\Psi_\beta\right)\ .
\label{ss2}
\end{equation}
\textbf{Step 3}:  Similarly, focusing on $D3$ and noting that the $\downarrow$-branches of $\Psi_\beta$ are the $\spadesuit$-branches yields
\begin{equation}
P\left(\spadesuit\middle|\Psi_\alpha\right)=P\left(\downarrow\middle|\Psi_\beta\right)\ .
\label{ss3}
\end{equation}
\textbf{Step 4}:  Combining \eqref{ss1} - \eqref{ss3} gives
\begin{equation}
P\left(\diamondsuit\middle|\Psi_\alpha\right)=P\left(\spadesuit\middle|\Psi_\alpha\right)=P\left(\downarrow\middle|\Psi_\alpha\right)\ .
\end{equation}
Since these are all of the possibilities, the probability of each is a third and 
\begin{align}
P\left(\downarrow\middle|\Psi_\alpha\right)&=\frac{1}{3}
\nonumber
\\
P\left(\uparrow\middle|\Psi_\alpha\right)&=\frac{2}{3}\ .
\end{align}
This result holds whenever the reduced density matrix is as in \eqref{23reduced}.

The logic of this section suggests a way of thinking about the Born rule at an intuitive level. Our recipe amounts to the following prescription: write the state vector as a sum of orthogonal vectors with equal amplitudes by unitarily transforming the environment. Then (and only then), \emph{ESP} justifies according equal credence to each such basis vector.  In that sense, the Born rule is simply a matter of counting.  However, we don't need to take this picture literally.

Now that \emph{ESP-QM} has revealed its power, one might reasonably suspect that we have given it too much.  Let's take a moment to reflect on the principle's plausibility.  As was discussed in \textsection \ref{response} and \ref{sec21}, if \emph{ESP-QM} is correct then Alice's credence that the result was up should not change between $t_2$ and $t_3$ in \textsc{Once-Or-Twice} \emph{even though} the number of copies of Alice changes.  The proof for the $\frac{1}{3}$/$\frac{2}{3}$ case above relies on the same trick: the reduced density matrix in \eqref{23reduced} could describe a case of three detectors with three different combinations of outputs---as in \eqref{alphaandbetastates}---or a single detector with two different outputs.  By \emph{ESP-QM} the probability of an up result must be the same in either case.  When \emph{ESP} was introduced it was immediately restricted to only apply in cases where the changes made to the environment do not involve the creation of additional copies of the agent elsewhere.  Why?  We cannot claim that the problem is that such changes increase the number of copies in existence---the change from $t_2$ to $t_3$ in \textsc{Once-Or-Twice} does too.  One problem {with omitting the restriction is that it would be} impossible to assign non-zero credences to the additional copies in the environment, since the agent {would have to} assign the same probabilities to all of the original copies whether or not additional copies are present.  This problem does not arise when \emph{ESP-QM} is applied to the quantum case.  The change from $t_2$ to $t_3$ in \textsc{Once-Or-Twice} increases the number of copies of Alice in existence, but she is not forced to assign any of the copies probability zero.  The fact that \emph{ESP-QM} avoids this particular problem does not fully exonerate the principle.  The change in the quantum state from $t_2$ to $t_3$ \emph{could} conceivably turn out to be relevant.  The motivation for \emph{ESP-QM}, emphasized earlier, is that when Alice is wondering about her relation to some detector $D$, things happening elsewhere are \emph{in fact not} relevant.  We believe the principle to be well-motivated but not established beyond any doubt, and thus our derivation is provisional: if one can focus on the reduced density matrix in calculating probabilities as recommended by \emph{ESP-QM}, the Born rule follows.

\section{Varieties of Uncertainty}\label{varieties}
\subsection{ESP and Indifference}\label{ESPIndifference}

In classical physics, you experience self-locating uncertainty when, for example, the universe is so large that you should expect there to exist a distant planet where someone is having the exact same immediate experiences that you are, or when the universe survives so long that you should expect short-lived Boltzmann brains to pop out of the vacuum in the exact subjective state you are in now, or when, as in \textsc{Duplicating Dr. Evil}, you have reason to think someone has purposefully created a duplicate of you.  Such phenomena can also occur in quantum mechanical contexts, leading to within-branch uncertainty.  Once the principle is strengthened, \emph{ESP} mandates that this sort of uncertainty should be treated with \emph{Indifference}.  This brings out an important virtue of the epistemic principle we've proposed: \emph{ESP} explains why \emph{Indifference} was a good heuristic valid in a wide variety of cases and also explains why the uncertainty arising from quantum measurements should be treated differently, using the Born rule.\footnote{{In a recent paper, \citep{wilson2015} has offered a clever alternative explanation as to why within-branch uncertainty should be treated differently from which-branch uncertainty: different Everettian worlds are in fact different possible worlds and thus \emph{Indifference} does not constrain one's treatment of which-branch uncertainty.  Our analysis goes further than just explaining the difference between the two kinds of uncertainty, we also explain the commonality (by giving a unified account of how both the Born rule, quantifying which-branch uncertainty, and \emph{Indifference}, quantifying within-branch uncertainty, arise from a single core epistemic principle, \emph{ESP}).\label{wilsonmove}}}

To apply \emph{ESP} fruitfully to classical cases, it will help to strengthen the principle.  The stronger version of \emph{ESP} is not concise when stated precisely (below), but the basic idea is simple:  Not only should it be irrelevant what's going on outside of the subsystems in $\mathcal{S}$, it also should be irrelevant where in spacetime each subsystem in $\mathcal{S}$ is located.  Consider again \textsc{Duplicating Dr. Evil}.  It seems irrelevant where the philosophers decide to build the replica of the lab.  They could make it in America or Japan or on Mars.  The choice shouldn't affect the probability that Dr. Evil assigns to being on the moon.  Perhaps more controversially, we believe it is irrelevant \emph{when} they build the replica.  Suppose they tell Dr. Evil that they are scanning his lab now and will (unstoppably) make the replica next week.  In such a case, Dr. Evil should start to wonder whether he's mistaken about the date.  Further, we believe his doubts about being on the moon should not be mitigated by the temporal separation between himself and the replica.
\begin{description}
\item[Strong ESP]  
Suppose that universe $U$ contains within it a set of subsystems $\mathcal{S}$ such that every agent in an internally qualitatively identical state to agent $A$ is located in some subsystem which is an element of $\mathcal{S}=\{X,Y,...\}$.  Let $\mathcal{S}'=\{M_X(X),M_Y(Y),...\}$ where each $M(\cdot)$ is a transformation which rotates, spatially translates, and/or temporally shifts\footnote{\emph{Strong ESP} loses some of its plausibility if these three transformations are not symmetries of the dynamical laws of $U$ and $U'$.  Here we assume that they all are.} the subsystem.  The probability that $A$ ought to assign to being located in a particular subsystem $X\in\mathcal{S}$ given that they are in $U$ is identical to the probability that they ought to assign to being in $M_X(X)\in\mathcal{S}'$ given that they are in some universe $U'$ (which contains within it a set of subsystems $\mathcal{S}'$ such that every agent in an internally qualitatively identical state to agent $A$ is located in some subsystem which is an element of $\mathcal{S}'$).
\begin{equation}
P(X|U)=P(M_X(X)|U')
\label{sesp}
\end{equation}
\end{description}
Here $M_X(X)$ is a {relocated} version of $X$, $M_Y(Y)$ is a {relocated} version of $Y$, etc.  Since $M_Y(Y)$ is simply a {relocated} version of subsystem $Y$, being in $M_Y(Y)$ will feel just like being in $Y$.  One might worry that the fact that $M_Y(Y)$ is surrounded by a different local environment than that around $Y$ could make $M_Y(Y)$ and $Y$ distinguishable.  If the replica of the laboratory is built floating in outer space instead of sitting on Earth's surface, the instantaneous arrangement of furniture and fermions may be the same, but the copy of Dr. Evil would quickly notice the absence of a force keeping his feet on the floor.  Still, if we focus on the \emph{instant}\footnote{If one thinks (reasonably enough) that it takes time to have an experience, the subsystems $X$, $Y$, ... must be taken to be temporally extended if they are to contain agents having experiences.  For example, let $Y$ be the terrestrial laboratory persisting over the course of, say, one minute.  {Moving} $Y$ into outer space wouldn't change what's happening inside the laboratory during that minute at all. It would, however, involve a violation of the laws of nature (as the felt downward force would have no source) unless other appropriate changes were made (like the addition of an appropriately sized planet beneath the laboratory's floor).\label{temporalextension}} when the outer space replica really is a perfectly shifted version of the terrestrial replica, all of the particles in the each agent's brain are in the same arrangement and the outer space copy must be having exactly the same experiences as those had by the terrestrial copy (and thus the same experiences as those of the original Dr. Evil).

Using \emph{Strong ESP}, it is straightforward to prove that one must follow the recommendations of \emph{Indifference} in cases of classical duplication.  In fact, the proof is so quick that it may cause you to doubt the principle.  It shouldn't.  \emph{Indifference} was intuitively plausible.\footnote{In fact, it's not just intuitive, there are arguments for it.  The argument given by \citet{elga2004} essentially relies on something like \emph{ESP}, as we discussed in Section~\ref{sec21}.}  \emph{Strong ESP} retains those intuitive recommendations while avoiding the unacceptable recommendations in cases like \textsc{Once-or-Twice} (\textsection \ref{response}).

We'll first apply \emph{Strong ESP} to \textsc{Duplicating Dr. Evil} and then move to the general case.  Let $X$ be the lunar laboratory and all of its contents (including Dr. Evil) and $Y$ be the terrestrial replica and its contents.  Let $M_X(X)$ move the lunar laboratory to where the terrestrial replica was and $M_Y(Y)$ move the lunar laboratory to where the terrestrial replica was ($M_X(X)=Y$ \& $M_Y(Y)=X$, provided the two laboratories are in truly identical physical states).  Take $U$ to be the original world and $U'$ to be the universe you get by switching the two laboratories (which, as it happens, doesn't change anything: $U'=U$).  Then, \eqref{sesp} yields
\begin{equation}
P(X|U)=P(M_X(X)|U')=P(Y|U)\ .
\end{equation}
Generalizing to arbitrary many copies of an agent is simple.  By pairwise swaps like the one above it can be shown that any two copies are equiprobable.  Thus when we restrict attention to cases where the agent's copies are \emph{physically} identical and not merely having identical \emph{experiences}, the recommendations of \emph{Strong ESP} align with those of \emph{Indifference}.  The recommendations even agree in cases of quantum measurement when the agent's copies are truly identical, as in equal amplitude superpositions like \eqref{state15050}.

The above justification for assigning equal credence to being in the lunar and terrestrial laboratories in the \textsc{Duplicating Dr. Evil} case at no point relied on the laws being classical.  One could equally well apply the reasoning to cases of within-branch uncertainty.  For simplicity, focus on a single branch which contains a copy of Dr. Evil on the moon and another on Earth.  The state might then be represented as $|\Psi\rangle=|L\rangle |T\rangle |E\rangle$, where $|L\rangle$ is the lunar lab, $|T\rangle$ is the terrestrial lab, and $|E\rangle$ is everything else.  Take subsystem $X$ to be the lunar laboratory, represented by the reduced density matrix $|L\rangle\langle L |$, and $Y$ to be terrestrial one, $|T\rangle\langle T |$.  We can define two unitary shift operators: $\widehat{S}_X$ which takes $|L\rangle$ to $|T\rangle$ and $\widehat{S}_Y$ which takes $|T\rangle$ to $|L\rangle$.  With these operators we can express $M_X$ and $M_Y$ mathematically as $M_X(\cdot)=\widehat{S}_X(\cdot)\widehat{S}^\dagger_X$ and $M_Y(\cdot)=\widehat{S}_Y(\cdot)\widehat{S}^\dagger_Y$.  The remainder of the argument proceeds as above.  Note that the division into subsystems is different here than in \textsection \ref{prf}.  There we assumed you know what particles you are made of and are trying to determine what branch of a particular reduced density matrix you inhabit.  Here we've focused on the question of what you're made of---that is, which reduced density matrix gives the state of the particles that compose your body.

\subsection{Mixed Uncertainties}\label{mixed}

We have now seen that, depending on the scenario, probability in quantum mechanical contexts is sometimes handled by the Born rule and other times handled by \emph{Indifference}.  What about cases where the two types of uncertainty are mixed---when one knows neither where they are in spacetime nor which branch they are on?  Here we must rely on the rule from which both individual prescriptions arise: \emph{Strong ESP}.  In this section, we'll discuss two illustrative `quantum sleeping beauty' scenarios that combine both kinds of uncertainty (\citeauthor{plewis}, \citeyear{plewis}, \citeyear{lewis2009reply}; \citeauthor{papineau2009}, \citeyear{papineau2009}, \citeyear{papineau2009reply}; \citealp{peterson,bradley2011,bradley2014,wilson2013,groisman2013}).

\begin{description}[font=\normalfont\scshape]
\item[Two-Branch-Beauty] The experimental subject, Beauty, will be put to sleep on Sunday night.  While she is asleep, there will be a $z$-spin measurement of particle $a$ which is initially $x$-spin up.  If the result is up, she will be awoken on Monday and then her memory of Monday's events will be erased so that on Tuesday when she wakes up her last memories will be of going to sleep Sunday night.  If the result is down, she will not have her memory erased.  Beauty knows everything that might be relevant about the setup.
\end{description}

\begin{figure}[htb]
\center{\includegraphics[width=14cm]{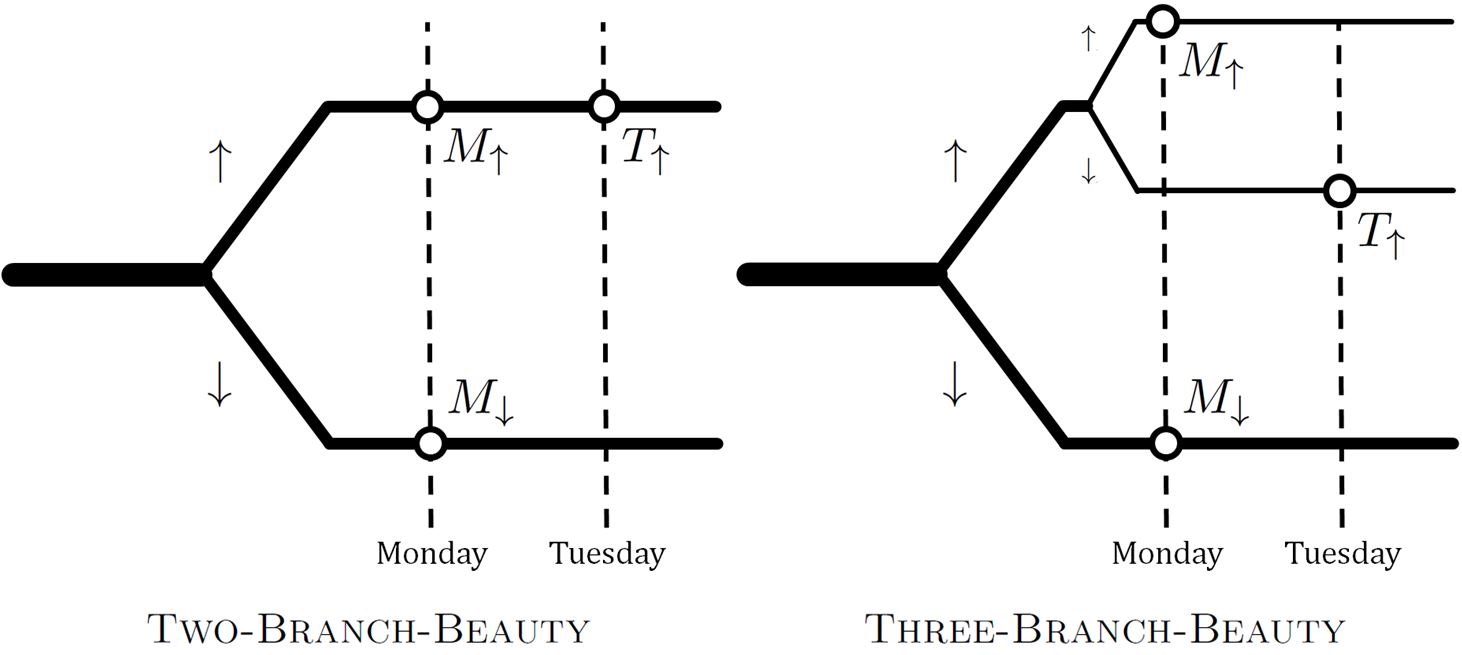}}
\caption{Two quantum sleeping beauty scenarios.}
\label{}
\end{figure}

Consider {Beauty's} situation upon waking with her most recent memories being those of going to bed Sunday evening. {Her} evidence doesn't discern between three possible locations in the multiverse where she might be: Monday morning on a branch where $a$ was measured to be up, $M_\uparrow $, Tuesday morning on a branch where $a$ was up, $T_\uparrow$, or Monday morning on a branch where $a$ was down, $M_\downarrow$.  Upon waking, what probability should Beauty assign to particle $a$ being up?  That is, what probability should she assign to $M_\uparrow \vee\: T_\uparrow$?  By \emph{ESP}-based arguments like those in \textsection \ref{prf}, she should assign equal probability to $M_\uparrow$ and $M_\downarrow$,
\begin{equation}
P\left(M_\uparrow\right)=P\left(M_\downarrow\right)\ .
\label{partone}
\end{equation}
However, since $M_\uparrow$ and $M_\downarrow$ are not the only alternatives, it does not follow that she should assign a probability of one half to each.

Given that $a$ was spin up, we're dealing with a case of within-branch uncertainty about what day it is.  Here \emph{Indifference} and \emph{Strong ESP} agree that Alice should assign equal probability to it being Monday or Tuesday morning (provided that the Monday and Tuesday Alices are in identical physical states, a useful albeit unrealistic assumption), 
\begin{equation}
P\left(M_\uparrow\right)=P\left(T_\uparrow\right)\ .
\label{parttwo}
\end{equation}
Since $M_\uparrow$, $M_\downarrow$, and $T_\uparrow$ are the only three options, it follows from \eqref{partone} and \eqref{parttwo} that Alice should assign each alternative a probability of one third.  The answer to our original question is that $P\left(\uparrow\right)=P\left(M_\uparrow \vee\: T_\uparrow\right)$ ought to be two thirds.  We thus recover the popular `thirder' result for this kind of sleeping beauty scenario.\footnote{\citet[\textsection 6]{vaidman2011} gets the same result in this case through his consideration of self-locating uncertainty.}

\textsc{Two-Branch-Beauty} may appear to be merely a fanciful philosopher's concoction, of little to no physical importance.  In fact, the case is importantly similar to inflation scenarios where some quantum branches have no life, others have life, and some have so much life that one can reasonably expect that one's own experiences are duplicated somewhere else in spacetime.  If we hope to develop a framework for testing theories like this, we need epistemic principles which can handle uncertainty about both which branch one is on and where one is within that branch.

There's something odd about the recommendations of \emph{Strong ESP} in \textsc{Two-Branch-Beauty}.  The probability that the principle tells Beauty to assign to up is $2/3$ whereas the Born rule recommends a probability of $1/2$.  Is the Born rule invalid?\footnote{See the discussion of cosmological cases where it appears that the Born rule is untenable in \citep{Page:2009qe}; see also \textsection \ref{generalmethod}.}  We think that the right lesson to draw is that there are two sources of uncertainty here and the Born rule is perfectly correct for quantifying the uncertainty brought about by quantum measurement.  However, there is also uncertainty arising from the duplication of Beauty's experiences on the up branch.  When both types of uncertainty are present, we need a rule for aggregating them.  \emph{Strong ESP} is capable of doing this, as discussed here and in \textsection \ref{generalmethod}.

To understand what's special about the \textsc{Two-Branch-Beauty} case discussed above, it is worth considering a variant of the case.
\begin{description}[font=\normalfont\scshape]
\item[Three-Branch-Beauty] Beauty will be put to sleep on Sunday night.  While she is asleep, there will be a $z$-spin measurement of particle $a$ which is initially $x$-spin up.  If the result is up, particle $b$ (identically prepared) will also be measured: If $b$ is up, Beauty will awake on Monday morning; if down, Beauty will be kept asleep until Tuesday morning.  If particle $a$ is measured to be down, particle $b$ will not be measured and Beauty will awake on Monday.  Beauty knows the setup.
\end{description}
This is simply a case of repeated quantum measurements, essentially the same setup as \textsc{Once-or-Twice}.  Here we can run arguments like those in \textsection \ref{prf} to show that Beauty should assign Born rule probabilities: $P\left(M_\uparrow\right)=1/4$, $P\left(T_\uparrow\right)=1/4$, and $P\left(M_\downarrow\right)=1/2$.\footnote{A wrinkle:  Before we can apply the methods of \textsection \ref{prf}, one must first shift attention to a universe where the $T_\uparrow$ copy of Beauty is temporally translated back to Monday.  \emph{Strong ESP} says that the probabilities will be the same in such a universe.  However, this new case is easier to handle as it is exactly analogous to \textsc{Once-or-Twice}.}  Comparing \textsc{Three-Branch-Beauty} and \textsc{Two-Branch-Beauty}, we see that \emph{Strong ESP} explains how the popular thirder solution can be correct for \textsc{Two-Branch-Beauty} whereas the halfer solution is correct in the quantum \textsc{Three-Branch-Beauty} case (solving the mystery of \citealp{plewis}).

\subsection{Large Universe Cosmology and the Quantum Multiverse}\label{generalmethod}

The tricks of \textsection \ref{mixed} can be used to give recommendations in the general case.  Consider a universe $U$---recall that in our terminology `universe' means `entire quantum multiverse'---with a set of observers $\{\calo_i\}$ who find themselves in indistinguishable circumstances.\footnote{Here we work under the fiction that the observers can be easily distinguished, labeled, and counted.  In actuality there may be multiple ways of carving the wave function into branches {(see \textsection \ref{sluEM})}.  Corresponding to different ways of carving the wave function into branches, there will be different collections of observers $\calo_i$.  However, predictions derived using \eqref{palpha} will not depend significantly on the particular carving.  In applying \emph{Strong ESP} we also assume that the observers are physically identical (modulo the weights of their branches).} They may be on the same branch of the wave function, on different branches, or even at different times. For each observer $\calo_i$ existing at some time $t_i$, the overall state describing the universe can be written in the form
\begin{equation}
  |\Psi(t_i)\rangle = \alpha_{i} |\phi_i\rangle + \beta_i|\phi^\perp_i\rangle\ ,
\end{equation}
where $|\phi_i\rangle$ is the branch on which the observer lives and $|\phi^\perp_i\rangle$ is the remainder of the quantum state, including all other branches.  $|\phi^\perp_i\rangle$ is (at least approximately) orthogonal to $|\phi_i\rangle$.

In this case, by the considerations above, \emph{Strong ESP} provides an unambiguous procedure for assigning credences. To each observer we assign a weight
\begin{equation}
  w_i = |\alpha_i|^2\ .
\end{equation}
Then the probability for being observer $\calo_i$ is simply\footnote{\citet{groisman2013} have proposed essentially the same rule.}
\begin{equation}
  P(\calo_i|U) = \frac{w_i}{\sum_j w_j}\ .
  \label{palpha}
\end{equation}
Note that the weights $\{w_i\}$ will not in general sum to unity for a variety of reasons: the weights may be calculated at different times for different observers; there may be multiple observers on a single branch; there may be branches with no observers. This rule reduces to {\it Indifference} in the case of multiple observers on a single branch and to the Born rule when there is a single observer on each branch.

This recipe provides a resolution of the ambiguity in applying the Born rule in large universes that was identified by \citeauthor{Page:2009qe} (\citeyear{Page:2009qe}, \citeyear{Page:2009mb}, \citeyear{Page:2010bj}); see also \citep{Aguirre:2010rw,Albrecht:2012zp}. Consider a universe that is large enough to contain multiple observers  with identical experiences on the same branch of the wave function. Imagine that each observer plans to measure the $z$-spin of their particle, each particle being in a potentially different pre-measurement state:
\begin{equation}
  |\sigma_i\rangle = \gamma_i|\uparrow_z\rangle + \delta_i|\downarrow_z\rangle\ .
  \label{singlespin}
\end{equation}
In simple cases of quantum measurement, the Born rule can be expressed as the statement that the probability of an observational outcome is given by the expectation value of a projection operator. For example, in an individual state of the form (\ref{singlespin}), the probability of observing spin-up is $P({\uparrow_z}|\sigma_i) = \langle\sigma_i|\hat{\Pi}_\uparrow|\sigma_i\rangle = |\gamma_i|^2$, where $\hat{\Pi}_\uparrow = |{\uparrow_z}\rangle\langle {\uparrow_z}|$. Page shows that there is no projection operator that gives the probability that an observer, not knowing which observer they are, will measure $\uparrow_z$ or $\downarrow_z$; in that sense the Born rule is insufficient to fix the probabilities of measurement outcomes. \emph{Strong ESP} resolves this ambiguity; using (\ref{palpha}), the probability of $z$-spin up (post-measurement, pre-observation) is simply given by
\begin{equation}
  P({\uparrow_z}|U) = \sum_i P(\calo_i|U)\times|\gamma_i|^2\ ,
\end{equation}
and analogously for $z$-spin down.  Here $P(\calo_i|U)$ are the probabilities assigned to being different observers \emph{before} the measurement is conducted.

This result has consequences for the measure problem in cosmology \citep{Freivogel:2011eg, Salem:2011qz, Vilenkin:2013ik}, although we will not explore them in detail here. We will only note that the probability one should assign to being a particular observer in the multiverse clearly depends on the amplitude of the branch on which that observer finds themselves. For example, consider one evolving branch on which the temporal density of observers grows exponentially, $n(t) \propto e^{\omega t}$, such as might happen in inflationary cosmology.  Let $\mathcal{B}$ be the subset of observers who live on that branch.  According to our prescription, the probability one should assign to being on that branch is not proportional to the integral of $n(t)$ over time, which is obviously infinite. Rather, it should be weighted by the amplitude $\alpha(t)$ of the corresponding branch:
\begin{equation}
  P(\mathcal{B}|U) \propto \int dt\, |\alpha(t)|^2 n(t). 
\end{equation}
If the the amplitude is decaying exponentially in time, $\alpha(t) \propto e^{-\gamma t}$, we will have a well-defined finite probability for being a member of $\mathcal{B}$ as long as $\gamma > \omega/2$.  If $\gamma < \omega/2$, the numerator and denominator of \eqref{palpha} both go to infinity and the rule fails to give a well-defined probability for being in $\mathcal{B}$.

\section{Probability in Practice}\label{pinp}

In \textsection \ref{prf} we showed that, post-measurement pre-observation, agents should assign probabilities to measurement outcomes in accordance with the Born rule.  This alone is not a sufficiently strong result to show that the many-worlds interpretation recovers all of the important aspects of quantum probability.  There remain two key problems, identified in \citet{papineau, greaves2007b}.  First, \emph{the practical problem}.  When faced with decisions whose repercussions depend upon the outcomes of future quantum measurements, why should agents act as they would if only one outcome were going to occur with probability determined by the Born rule?  Second, \emph{the epistemic problem}.  Why can we infer facts about the wave function from observed long-run frequencies?  Also, why do the data usually taken as evidence for quantum mechanics provide evidence for \emph{Everettian} quantum mechanics?  As we are primarily concerned with our reasons for believing in the many-worlds interpretation, we focus on the epistemic problem.  The practical problem is less urgent.  If Everettian quantum mechanics is well-confirmed by the evidence but turns out to recommend that we act differently, then we would do well to adjust the way we make decisions. In \textsection \ref{betandbranch} we suggest that perhaps there is no adjustment, that an Everettian can continue to act as if a single outcome will occur with probability given by the Born rule.

A common objection to Everettian quantum mechanics is that there is no way for probabilities to arise in a deterministic theory when the entire physical state and the laws are known.  Consider \textsc{Once-or-Twice}.  At $t_1$ before the measurements have been made, Alice knows what will happen.  She will have a successor who sees that particle $a$ was measured to be down and two who see up.  If she knows the universal wave function, there is nothing for her to be uncertain of.  Of course, as was discussed in \textsection \ref{sluEM}, she will experience self-locating uncertainty at $t_2$ and $t_3$.  But, at $t_1$ there is no uncertainty.  Still, it may be that Alice should act \textit{as if} she was uncertain (perhaps there is, as \citeauthor{vaidman2011} \citeyear{vaidman2011}, \citeyear{vaidman2014} would say, an `illusion of probability').  Regardless of how Alice is to act at $t_1$, when the data is collected and the theories are tested, between $t_3$ and $t_4$, the requisite uncertainty \emph{is} present and confirmation can proceed as usual.  Here we follow \citet{greaves2004} in admitting that Everettian quantum mechanics lacks pre-measurement subjective uncertainty but arguing that it is not thereby refuted.\footnote{Although we will not explore the option here, one could try to revive some notion of pre-measurement uncertainty which would be present even in cases where the wave function is known; strategies include conceiving of persons as four-dimensional worms, taking Everettian worlds to be diverging as opposed to overlapping, and/or trying to most charitably interpret words like `uncertainty' as uttered by agents who unknowingly reside in an Everettian multiverse \citep[see][ch. 7]{saunders2008,saunders2010b,wilson2012b,wallace2012}.  Our proof of the Born rule post-measurement could  potentially be combined with such an account to justify setting the pre-measurement probabilities in alignment with the Born rule.  However, it is not clear that true pre-measurement uncertainty is needed to solve the practical or epistemic problems.  Although there may be some way of making sense of pre-measurement probabilities, we hope to show that our approach does not rely on the success of such a program (here we adopt the strategy of \citealp[\textsection 1.3]{greaves2007b}).\label{pbuncertainty}}  The core problem of this section, whether post-measurement pre-observation probabilities are sufficient for solving the practical and epistemic problems, has been discussed by \citet{tappenden2011} and we are largely in agreement with his conclusions.

\subsection{Betting and Branching}\label{betandbranch}

Imagine that, in \textsc{Once-or-Twice}, at $t_1$ Alice is offered a bet which costs $\$20$ and pays $\$50$ if the {first measurement yields} down, nothing if up. The net reward or cost is assessed at $t_4$.  Should Alice accept the bet?  At $t_2$, all of her successors will wish that she had taken the bet.  They each assign a probability of $0.5$ to up and $0.5$ to down, so the expected value of the bet is $\$5$.  At $t_3$, the expected value of the bet is the same and again all of Alice's successors will wish she had taken it.  Alice knows ahead of time that although she'll gain no new information about the outcome as time progresses, once the measurement has occurred she will wish she'd taken the bet.  So, it seems reasonable that at $t_1$ she should {gamble} in the way her future selves will wish she had and accept the bet. Generalizing this reasoning, Alice should {always} bet as if one outcome were going to occur with probability given by the Born rule since during the inevitable post-measurement pre-observation period all of her successors will approve of choices made under this supposition. (This type of argument is presented and assessed in \citealp[][]{tappenden2011}.)

When Alice decides, pre-measurement, to accept the bet, she is not doing so because she is trying to make a decision under uncertainty.  Instead, she is trying to make a decision about how to distribute goods among her successors.  If she distributes goods as recommended above, each successor will think the decision reasonable before they come to know which successor they are.  Once they come to realize which branch they're on, they may well not endorse the bet.  This should not be surprising.  Even the most careful gambler can lose and in such cases would prefer that the bet was never made.

The strategy for arriving at effective pre-measurement uncertainty outlined above has three distinct shortcomings.  First, the claim that one's pre-measurement decisions must align with the post-measurement preferences of one's successors must be justified.  This alignment might be enforced by some sort of decision-theoretic reflection principle, but such a principle would need to be precisely stated and defended.\footnote{Some work has already been done in this direction.  \citet[\textsection 8.1]{wallace2002} introduces a decision-theoretic reflection principle which is discussed in \citet[\textsection 4.2]{greaves2004} and \citet[\textsection 5.2.1]{greaves2007b}.  The requirement is roughly that one's betting behavior ought not change over a period of time in which no new evidence is gained (and this is coupled with the idea that the occurrence of an expected branching event provides no new evidence).  \citeauthor{wallace2010b} (\citeyear{wallace2010b},\citeyear{wallace2012}) defends a requirement of diachronic consistency which does similar work.  Unfortunately, both of these principles condemn agents we take to be acting rationally in various cases where self-locating uncertainty is important (such cases often violate \emph{epistemic} reflection principles, see \citealp{lewis2009, arntzenius2003}).  For example, the principles condemn an agent who accepts the \textsc{Duplicating Dr. Evil} Dutch book in appendix \ref{notaproblem} as the agent's betting behavior changes once the duplicate is created.  We thus believe that these two principles cannot be correct as stated.}  Second, this strategy only works if the preferences of agents are narrow: each successor only cares about what's happening in their own branch.\footnote{\citet{tappenden2011} is aware of this problem (which was raised in the context of anticipated branching by \citealp{price2010}).}  If each of Alice's successors only cares about their own wealth, then the bet described above seems lucrative.  There is a 50\% chance of winning \$30 and a 50\% chance of losing \$20.  The probability of being on a certain branch acts, for all practical purposes, like the probability that a certain thing is happening.  But, if the successors care about, say, the average wealth of all successors, then at $t_2$ it seems like a good bet but at $t_3$ it does not.  (Here we've assumed an idealized case, as in \textsection \ref{quant}, where the number of successors is well-defined.  Removing the idealization makes decision making harder, but doesn't fix the problem that one's preferences may extend beyond the goings on in one particular branch.  For example, one might conceivably desire that no copy of oneself elsewhere in the multiverse be experiencing a truly miserable life \citep[\textsection 7]{price2010}.)  Third, the strategy outlined above does not directly address cases where there is no period of uncertainty at all because observation happens immediately upon measurement. We believe such cases are atypical, but perhaps possible (\textsection \ref{sluEM}).  In such a case, one can use the strategy above to argue that \emph{had there been} a period of uncertainty of any length at all, it would have been rational to treat the psi-squared branch weights as probabilities for the sake of decision making---to care about one's successors in proportion to the weights of the branches they occupy.  As the limit of the amount of time after measurement before observation is taken to zero, the same decisions remain rational.  It would be very odd if they were not the right actions to take when there is no period of uncertainty at all.\footnote{A similar limiting argument is made by \citet[\textsection 5.2.1]{greaves2007b}, and another {appealing to} a hypothetical period of uncertainty by \citet[][\textsection 4]{tappenden2011}}

\subsection{Theory Confirmation}\label{inflink}

For the purposes of empirically testing competing theories---about the state of the system or the laws that govern its evolution---it is necessary that when the data come in we can judge whether the data are considered probable or improbable by the various theories under consideration.  That is, immediately before the agent \emph{observes} the result, probabilities of different outcomes must be well-defined.  However, it is inessential that there be well-defined probabilities before the \emph{measurement} is made and the outcome is recorded by the measuring device (the point at which the wave function branches).  Post-measurement pre-observation probabilities are sufficient for theory confirmation.  On our account, immediately before looking at the outcome of an experiment an agent in an Everettian multiverse is uncertain of what the observed outcome will be and is perfectly capable of quantifying that uncertainty.  Theories can be tested in familiar ways.

Although theory testing in Everettian QM proceeds essentially as usual, we discuss two kinds of learning scenarios here for the sake of illustration. First, consider testing theories about the wave function of the system by gathering data about which eigenvalues are measured.  We will treat the problem from a Bayesian perspective, but one could {apply alternative} methods \citep[see][\textsection 6.2 \& 6.3]{wallace2012}.  For simplicity, consider a case where there are just two theories under consideration:

\begin{description}[font=\normalfont\scshape]
\item[What Wave Function?] Alice will measure the $z$-spin of a single particle which she knows to be prepared in either state `$P\!\uparrow$' (`probably up') or state `$P\!\downarrow$,'
\begin{align}
|\Psi_{P\uparrow}\rangle &= \sqrt{\frac{9}{10}}\:\left|\uparrow_z\right\rangle+\sqrt{\frac{1}{10}}\:\left|\downarrow_z\right\rangle
\nonumber
\\
|\Psi_{P\downarrow}\rangle &= \sqrt{\frac{1}{10}}\:\left|\uparrow_z\right\rangle+\sqrt{\frac{9}{10}}\:\left|\downarrow_z\right\rangle\ .
\label{whatwf}
\end{align}
Before the experiment, she considers either wave function equally likely.  Let $t_1$ be a time after preparation before measurement, $t_2$ be post-measurement pre-observation, and $t_3$ be immediately post-observation.
\end{description}
On our account it is not hard to see that observing an up result confirms $\Psi_{P\uparrow}$.  At $t_2$, Alice assigns conditional probabilities $P_{t_2}\left(\uparrow_z\middle|\Psi_{P\uparrow}\right)=0.9$ and $P_{t_2}\left(\uparrow_z\middle|\Psi_{P\downarrow}\right)=0.1$.  As she knows branching has occurred, these are the probabilities she assigns to being on an up branch conditional on particular initial wave functions.  If upon observation she sees that the spin was up, she should update her credence in $\Psi_{P\uparrow}$ by conditionalizing on her new evidence,
\begin{align}
P_{t_3}\left(\Psi_{P\uparrow}\right)&=\frac{P_{t_2}\left(\uparrow_z\middle|\Psi_{P\uparrow}\right)P_{t_2}\left(\Psi_{P\uparrow}\right)}{P_{t_2}\left(\uparrow_z\middle|\Psi_{P\uparrow}\right)P_{t_2}\left(\Psi_{P\uparrow}\right)+P_{t_2}\left(\uparrow_z\middle|\Psi_{P\downarrow}\right)P_{t_2}\left(\Psi_{P\downarrow}\right)}
\nonumber
\\
&=\frac{0.9 \times 0.5}{0.9 \times 0.5+0.1 \times 0.5}=0.9\ .
\label{updating}
\end{align}
The credence Alice assigns to $\Psi_{P\uparrow}$ jumps from $0.5$ to $0.9$, as it should.  Continuing to observe more up results than down in identically prepared systems would further confirm the $\Psi_{P\uparrow}$ hypothesis over $\Psi_{P\downarrow}$.   In analyzing this case we've assumed that Alice should update as usual, by \eqref{updating}, even though some of the probabilities involved quantify her self-locating uncertainty, like $P_{t_2}\left(\uparrow_z\middle|\Psi_{P\uparrow}\right)$.\footnote{There is a way in which the case of theory testing here is different from normal cases.  On either of the two hypotheses about the wave function, it was guaranteed that a version of Alice would see spin up.  This leads to a worry:  What Alice learns upon seeing up is that `in my universe, one of the copies saw up'---which doesn't provide evidence for either theory over the other as Alice already knew with certainty that it would happen---and that `I'm the one who saw up'---which is purely self-locating information about where she is in the universe and thus not informative about the way the universe is.  We {side} with \citet{titelbaum2008} and \citet[pp. 294-295]{greavesM} {in rejecting the final step,} `purely self-locating evidence' {\emph{can be} informative about the way the world is}.  For a theory about learning from evidence that disagrees on this point, see \citet{meacham2008}.}

Next, consider a second type of learning scenario.  We can empirically confirm Everettian quantum mechanics over competing empirically inequivalent hypotheses about the physical laws by observing Born rule statistics.  Suppose we run a large number of experiments without observing the outcomes (one could equally well look after each experiment, but this way things will be a bit simpler).  Some sequences of results would be deemed likely by the Born rule and others would be judged very improbable.  Suppose that the actual sequence of outcomes, $\mathscr{S}$, is one that the Born rule judges likely. {For example, imagine} that of a large number of measurements of $\widehat{A}$ on systems prepared in the same state $|\Psi\rangle$, the fraction in which eigenvalue $a$ is observed  is approximately $|\langle a|\Psi\rangle|^2$.  Then, the probability of the evidence given the theory, $P(\mathscr{S}|\text{EQM}\&\Psi)$, is high for Everettian quantum mechanics, EQM, because Everettian probabilities match the Born rule probabilities (\textsection \ref{prf}).  If another theory considered such sequences to be less probable, then updating on $\mathscr{S}$ would support Everettian quantum mechanics over this competing theory.  If the alternative theory assigned the same probability to $\mathscr{S}$, the data would not discern between the two theories.

In general, confirmation will work as usual in cases where Alice has a period of uncertainty before the outcome is revealed to her.  What happens if in {\sc What Wave Function?} Alice sees that the particle is $\uparrow_z$ \emph{exactly when} the branching occurs so that there is no post-measurement pre-observation period?  This is a tricky question.  Alice might try to update her credences immediately after branching by conditionalization as in \eqref{updating} (now moving directly from the pre-measurement time $t_1$ to the post-observation time $t_3$ as there is no post-measurement pre-observation period),
\begin{equation}
P_{t_3}\left(\Psi_{P\uparrow}\right)=\frac{P_{t_1}\left(\uparrow_z\middle|\Psi_{P\uparrow}\right)P_{t_1}\left(\Psi_{P\uparrow}\right)}{P_{t_1}\left(\uparrow_z\middle|\Psi_{P\uparrow}\right)P_{t_1}\left(\Psi_{P\uparrow}\right)+P_{t_1}\left(\uparrow_z\middle|\Psi_{P\downarrow}\right)P_{t_1}\left(\Psi_{P\downarrow}\right)}\ .
\label{}
\end{equation}
In this equation, the probability $P_{t_1}\left(\uparrow_z\middle|\Psi_{P\uparrow}\right)$ is hard to interpret.  It's definitely not the probability that the particle was up given $\Psi_{P\uparrow}$ \emph{at} $t_1$ since the particle was in a superposition of up and down before the measurement was made.  It cannot be Alice's pre-measurement probability that `\emph{I} will see up' given that the particle is currently in state $\Psi_{P\uparrow}$ since the claim is ill-defined; some successors will see up and others will see down, all are Alice's descendants.  It should not be interpreted as the probability that \emph{some} successor of Alice sees up.  In cases where stochastic theories are under consideration this proposal would make Everettian QM too easy to confirm---all outcomes that one might observe are given probability one \citep[see][p. 140]{greaves2007b}.  {Standard Bayesian methods break down.}  The question of how to revise Bayesian confirmation theory is difficult and numerous proposals have been made---many of which rely on Elga's indifference principle and thus are incompatible with \emph{ESP} (see footnote \ref{listofproposals}).  To thoroughly address the question of how to update when the outcome is immediately observed upon branching, we must await philosophical progress.\footnote{One appealing recent suggestion is to replace $P_{t_1}\left(\uparrow_z\middle|\Psi_{P\uparrow}\right)$ with $\tau\!\left(\Psi_{P\uparrow}\rightarrow\:\uparrow_z\right)$ where $\tau\!\left(\Psi_{P\uparrow}\rightarrow\:\uparrow_z\right)$ is a `transition probability,' capturing something like the degree to which the copies of Alice on $\uparrow_z$ branches are Alice's successors if the initial wave function is $\Psi_{P\uparrow}$; or, more operationally, specifying how much of Alice's credence in $\Psi_{P\uparrow}$ should shift to the $\uparrow_z$ branches at $t_3$ before she takes account of any new evidence \citep{schwarz}.  If Schwarz's suggestion is adopted, one could argue that the transition probabilities should be given by the amplitude-squared of the wave function by requiring that belief update when there is no period self-locating uncertainty agree with the case of an arbitrarily small period of self-locating uncertainty.} {There is reason} to be optimistic.  {Plausibly,} in the case of immediate observation one should adjust their beliefs in the way they would \emph{if there were} a short period of self-locating uncertainty as the adjustment is the same no matter how short the period of self-locating uncertainty is \citep[p. 107]{tappenden2011}.

\section{Comparison to Other Approaches}\label{review}

The main purpose of this article is to present our derivation of the Born rule.  It is consistent with the success of our derivation that other attempts to derive the Born rule are also satisfactory. However, there would have been little motivation to embark on this project if we did not have concerns about existing approaches. In this section we will highlight the differences between our derivation and two existing programs: Zurek's envariance-based derivation and the decision-theoretic approach.

\subsection{Zurek's Envariance-based Derivation of the Born Rule}

Here we will briefly present \citetalias{zurek2005} \citeyearpar{zurek2005} argument in the simplest case and mention some of the limitations of his approach.

According to Zurek, the probabilities for a system $S$ to manifest various properties upon measurement depend only on the state of the system (`Fact 2').  Further, the state of a system is not affected by unitary transformations on the environment (`Fact 1').  From these assumptions, one can prove that the probabilities for different measurement outcomes are in agreement with the Born rule in ordinary scenarios.  For example, suppose the state of the universe is 
\begin{equation}
\frac{1}{\sqrt{2}}\Big(\ket{\uparrow_z}_S\ket{E_1}+\ket{\downarrow_z}_S\ket{E_2}\Big)\ ,
\end{equation}
where $E_1$ and $E_2$ are orthogonal.  It {is taken to follow} from the perfect entanglement between system and environment that $P(\uparrow_z)=P(E_1)$.  The following universal state can be reached by a unitary swap of the environment states (or a unitary swap of the system states), 
\begin{equation}
\frac{1}{\sqrt{2}}\Big(\ket{\downarrow_z}_S\ket{E_1}+\ket{\uparrow_z}_S\ket{E_2}\Big)\ .
\label{swapped}
\end{equation}
This swap does not change the state of the system or the environment, it only affects the entanglement between the two.  The perfect entanglement of \eqref{swapped} yields $P(\downarrow_z)=P(E_1)$.  Combining these two results gives $P(\uparrow_z)=P(\downarrow_z)$, in agreement with the Born rule.\footnote{In our exposition of \citep[\textsection II.C]{zurek2005} we take option 3 since it seems strongest and is most similar to the reasoning in our derivation.}

Zurek has provided a compelling argument that the Born Rule gives the only \emph{sensible} probability measure in Everettian quantum mechanics. However, that has arguably already been established by Gleason's Theorem \citep{gleason}.  What Zurek fails to explain is how probabilities arise at all in this deterministic theory---self-locating uncertainty is not discussed.  Thus in his derivation the nature of the probabilities involved is obscure.  Zurek claims that he is calculating probabilities for \textit{future} measurement outcomes, but he does not explain clearly what these are probabilities \textit{of}, saying that the `observer can be ignorant of his future state, of the outcome of the measurement he has decided to carry out' \citep[][\textsection VII.C]{zurek2005}.\footnote{See also FAQ \#5 of \citep{zurek2010}.}  This would be reasonable if only one outcome were expected to occur.  However, as explained in \textsection \ref{sluEM}, it is not obvious that there is anything for one to assign probabilities to in the many-worlds interpretation when all outcomes will occur and the future evolution of the wave function is known.\footnote{\citet[\textsection III.C]{zurek2005} recognizes this challenge and responds by departing from the many-worlds interpretation and discussing his Existential interpretation instead.  This raises doubts about the applicability of his proof to Everettian quantum mechanics.

\citet[ch. 7, esp. \textsection 7.6]{wallace2012} argues that we should expand our concept of subjective probability so that one can not only assign probabilities to ways the world might be and locations they might have in it, but also to different futures within a branching Everettian multiverse.  Doing this would help preserve the correctness of language used by agents in Everettian worlds.  Wallace may be right that such an extension is possible and advisable, but to the extent that it is, it requires a significant change in the way we understand probability.  In this new context, fact 2 is far from obvious.  Wallace's derivation of the Born rule is not susceptible to similar concerns as his assumptions are about the \emph{preferences} of ideal agents, not their \emph{probability assignments}.}  One could reasonably argue that Facts 1 and 2 are insufficiently motivated in Zurek's treatment, especially given that the probabilities involved are not the usual sort.\footnote{See also (\citealp[esp. \textsection III.F2]{schlosshauer2005}; \citealp[\textsection 3.1]{albert2010}).}  Although our {derivation} relies on an assumption as well, {\emph{ESP-QM}} (which is similar to Facts 1 and 2 taken together), we have attempted to provide a more thorough justification of our assumption and a more philosophically careful treatment of the probabilities involved.

\subsection{The Decision-theoretic Program}

Starting with \citep{deutsch1999}, there have been a variety of attempts to justify Born rule probabilities using decision theory (\citeauthor{greaves2004}, \citeyear{greaves2004}, \citeyear{greaves2007b}; \citealp{greavesM}; \citeauthor{wallace2003b}, \citeyear{wallace2003b}; \citeyear{wallace2010b}, \citeyear{wallace2012}; \citealp{wilson2013objective}).  The basic strategy is to argue that plausible constraints on rational preferences ensure that when we make bets about the outcomes of future quantum measurements we will act as if we assign Born rule probabilities to the various branches.  The success of their program is compatible with the success of our purely epistemic arguments.  In fact, the mathematical methods used to derive the Born rule are quite similar.\footnote{Compare \citetalias{wallace2012}, \citeyear[\textsection 5.5]{wallace2012}, use of erasure to ignore irrelevant details and our use of \emph{ESP-QM} in \textsection \ref{prf} to do the same.}  At this point, many remain unconvinced by the decision-theoretic program (\citealp{baker2007, albert2010, price2010, dizadji2013, maudlinreview}).  We focus on one particular type of concern here to highlight an advantage of our approach.

The constraints proposed in the decision-theoretic approaches may reasonably be doubted as they manifestly conflict with a \emph{prima facie} reasonable epistemic principle, \emph{Indifference}.  Deutsch's original formulation of the decision-theoretic argument for the Born rule implicitly appeals to the following assumption \citep{wallace2003b}:
\begin{quotation}
\noindent\small{\textbf{Measurement Neutrality.}  A rational agent is indifferent between any two quantum bets that agree on the state $\ket{\Psi}$  on which the measurement is to be performed, the observable $\widehat{X}$ to be measured, and the `payoff function' $P$ from the spectrum of $\widehat{X}$ to the set of consequences. \citep[p. 119]{greaves2007}}
\end{quotation}
\citet{wallace2007} has proved that \textit{Measurement Neutrality} is equivalent to:
\begin{quotation}
\noindent\small{\textbf{Equivalence.}  A rational agent is indifferent between any two quantum bets that agree, for each possible reward, on the mod-squared measure of branches on which that reward is given. \citep[p. 119]{greaves2007}}
\end{quotation}
The arguments in (\citealp{deutsch1999}; \citeauthor{greaves2004}, \citeyear{greaves2004}, \citeyear{greaves2007}; \citealp{wallace2003b}) appeal to this assumption in one of its manifestations. \emph{Measurement Neutrality} may have some intuitive plausibility but, looking at \emph{Equivalence}, it is obvious that this assumption will be inconsistent with a branch-counting rule for assigning probabilities.  The principle seems to beg the question against the defender of \emph{Indifference} who takes branch-counting to be the proper way to assign probabilities in Everettian quantum mechanics.  Defenders of the decision-theoretic program typically point out that the conflict with branch-counting should not cause doubt about \emph{Equivalence} since there are independent reasons to think branch-counting is wrong.  Two main reasons are given.  First, branches cannot generally be counted so branch-counting is not a rule one could actually apply.  As discussed in \textsection \ref{quant}, there are cases, albeit somewhat contrived, where branch number is well-defined and in these cases branch-counting gives definite recommendations which are in conflict with the Born rule. {The fact that} it is unclear how to apply a rule in some cases is not necessarily a reason to think it incorrect when it yields a clear judgment.  Second, branch-counting is diachronically inconsistent.  We do not take this to be a strong reason to reject branch-counting (see appendix \ref{notaproblem}).
 
In more recent decision-theoretic derivations of the Born rule, \citeauthor{wallace2010b} (\citeyear{wallace2010b}, \citeyear{wallace2012}) introduces a somewhat different collection of rationality axioms, multiple of which might be contested by the defender of branch-counting.  Wallace argues in favor of these axioms and does not take their conflict with branch-counting \citep[\textsection 5.8.1]{wallace2012} as a reason to think them incorrect, since branch-counting can be shown to be irrational.  For one who believes his axioms, this gives reason to doubt branch-counting.  But for the proponent of \emph{Indifference}, it gives reason to doubt Wallace's axioms. {We also reject \emph{Indifference} because it conflicts with an assumption about rationality (\emph{ESP}).  However, we take our argument to be potentially more persuasive to the proponent of \emph{Indifference} for two main reasons: \emph{ESP} is a single, simple, general epistemic principle which we believe has similar (if not more) initial plausibility than \emph{Indifference}; \emph{ESP} is consistent with \emph{Indifference} (and \emph{Strong ESP} entails it) in cases of classical self-locating uncertainty and thus it is capable of explaining why (and how) quantum cases are to be handled differently (though see footnote \ref{wilsonmove}).}  A strength of our approach, over both the decision-theoretic program and Zurek's derivation, is that it can be applied in both cases of classical duplication {and} in cases of quantum branching.

\section{Conclusion}

In this paper we have presented a justification for the Born rule in Everettian quantum mechanics in which self-locating uncertainty played a fundamental role. {The policy of reacting to self-locating uncertainty by treating each observer as equiprobable, \emph{Indifference}, is reasonable in classical scenarios but yields strange recommendations in quantum contexts.} Instead we proposed the \emph{Strong Epistemic Separability Principle} (\emph{Strong ESP}), which extends smoothly from \emph{Indifference} in the classical regime to the Born rule in cases of quantum measurement.

Following the consequences of \emph{Strong ESP} leads us to a simple and physically transparent derivation of the Born rule in the many-worlds interpretation. The appearance of probabilities in a deterministic theory is explained by evolution from perfect knowledge to unavoidable self-locating uncertainty. Our approach provides a unified perspective on uncertainties in both the classical and quantum contexts, with implications for large-universe cosmology as well as for the foundations of quantum mechanics.

There are a few ways in which future work could strengthen the approach to probability in Everettian quantum mechanics developed in this article.  First, although \emph{Strong ESP} has some intuitive appeal, one could seek further philosophical justification for the principle and a more precise formulation of it.  Second, more work could be done in justifying the appeal to reduced density matrices as the correct way of representing subsystems in Everettian quantum mechanics.  Third, one might hope to reach more definitive conclusions about the connection between post-measurement probabilities and pre-measurement decision making.  Fourth, the assumption in \textsection \ref{sluEM} that branching happens globally could be further defended in the context of a detailed account of when and how branching occurs.

\begin{center}
\textbf{Funding}
\end{center}

National Science Foundation (DGE 0718128 to C.S.);  Department of Energy (DE-FG02-92ER40701 to S.C.); Gordon and Betty Moore Foundation through Caltech Moore Center for Theoretical Cosmology and Physics (776 to S.C. and C.S.).

\begin{center}
\textbf{Acknowledgments}
\end{center}

Charles Sebens would like to thank John-Mark Allen, David Baker, Gordon Belot, Adrian Kent \citep[see][]{kent2014}, David Manley, Daniel Peterson, Barry Loewer, and Laura Ruetsche for very useful comments on drafts of this article.  He would also like to thank Simon Saunders and David Wallace for a wonderful introduction to the Everett interpretation.

Sean Carroll would like to thank Scott Aaronson, David Albert, Andreas Albrecht, Hirosi Ooguri, Simon Saunders, and David Wallace for helpful conversations.

\appendix

\section{What's Not Wrong with Branch-counting}\label{notaproblem}

\citet[\textsection 4.3]{wallace2012} gives the following diagnosis of why the switch from $P_{t_2}(\text{up})=\frac{1}{2}$ to $P_{t_3}(\text{up})=\frac{2}{3}$ in \textsc{Once-or-Twice} is irrational: it is diachronically inconsistent.\footnote{He credits Deutsch with giving a version of the argument in conversation \citep[][footnote 15]{wallace2013}.}  According to `the standard rules of the probability calculus,' the probability of particle $a$ being up at $t_3$ ought to be
\begin{equation}
P_{t_3}(\text{up at }t_3)=P_{t_2}(\text{up at }t_3|\text{up at }t_2)P_{t_2}(\text{up at }t_2)+P_{t_2}(\text{up at }t_3|\text{down at }t_2)P_{t_2}(\text{down at }t_2) \ .
\label{goodone}
\end{equation}
But, it clearly is not if
\begin{align}
P_{t_3}(\text{up at }t_3)&= \frac{2}{3}
\nonumber
\\
P_{t_2}(\text{up at }t_3|\text{up at }t_2)&= 1
\nonumber
\\
P_{t_2}(\text{up at }t_3|\text{down at }t_2)&= 0
\nonumber
\\
P_{t_2}(\text{up at }t_2)&= P_{t_2}(\text{down at }t_2)=\frac{1}{2}\ ,
\label{probs}
\end{align}
as recommended by \emph{Indifference}.  Without the subscripts, the point is certainly valid.  At $t_2$ or $t_3$ (or any other time), it must be the case that
\begin{equation}
P(\text{up at }t_3)=P(\text{up at }t_3|\text{up at }t_2)P(\text{up at }t_2)+P(\text{up at }t_3|\text{down at }t_2)P(\text{down at }t_2) \ ,
\end{equation}
provided `$\text{up at }t_2$' and `$\text{down at }t_2$' are mutually exclusive and exhaustive alternatives.  This forbids certain unsophisticated ways of filling in the branch-counting story, but is of no concern to the defender of \emph{Indifference}.  At $t_2$, $P(\text{up at }t_3)$ is only $1/2$. At $t_3$, $P(\text{up at }t_2)$ has risen to $2/3$.

\eqref{goodone} is not a requirement of \emph{consistency} for one's credences at a time, it is a constraint on the way one \emph{ought to} adjust their credences over time and not one that a proponent of \emph{Indifference} should accept.  At $t_2$ Alice knows that there's a $50/50$ chance that she's on an up branch and that if she is on one she'll definitely still be on one later.  However, once the second measurement is made she realizes there are three branches she might be on and that none of her evidence discerns between the three possibilities.  So, by the logic of \emph{Indifference}, she should think it more likely that she's on an up branch since there are twice as many of them.

\citet[\textsection 5.4]{wallace2012} rightly notes that because of the prevalence of branching in Everettian quantum mechanics, such credence shifts will be so common that deliberative action will be near impossible.  Such are the dangers of living in a wildly branching multiverse.

The violation of \eqref{goodone} can be made more worrisome as it leads to an Everettian Dutch book.\footnote{Wallace would likely see this as a friendly strengthening of his quick argument (see \citealp[\textsection 4.8]{wallace2012} together with \citealp[p. 247]{wallace2010xi}). For an alternate strategy, see \citep[\textsection II]{wallace2010xi}.}  If Alice reasons by \emph{Indifference}, there are a series of bets which can be given to her and her successors such that each bet will be judged fair but the combination of bets guarantees a loss on every branch.  Here is a way of constructing the Dutch book (from \citealp[\textsection 2]{peterson}):  At $t_2$ a bet is offered which pays $\$15$ if particle $a$ is down and $-\$15$ if up.  At $t_3$ a bet is offered which pays $\$10$ if particle $a$ is up and $-\$20$ if down.  All costs/rewards will be collected/paid at $t_4$.  If Alice assigns probabilities as in \eqref{probs}, these bets will seem fair.  However, if she accepts both bets then she will lose $\$5$ on each branch.  There is an apparent inconsistency here.  Looking forward from $t_1$ Alice can see that this series of bets will guarantee her a loss and thus would presumably choose not to take them if offered as a package before the branching starts.  However, as she goes through the experiment and her credences shift (in a way which she could have anticipated), she finds each bet fair when offered.

The defender of \emph{Indifference} should not be perturbed.\footnote{Thanks to David Manley for essential discussion here.  See \citep{lewis2009} for a related point.}  Consider this potential Dutch book for Dr. Evil in the \textsc{Duplicating Dr. Evil} case: Before $t$ (that is, before the duplication), Dr. Evil is offered a bet that some time long after $t$ will pay out $\$100$ to Dr. Evil and $-\$300$ to Dup.  Not long after $t$, he is offered a second bet which will pay out $\$200$ to Dup and $-\$200$ to Dr. Evil.  If both bets are accepted, Dr. Evil and Dup will each lose $\$100$.  The first bet seems lucrative.  Since Dup has yet to come into existence, Dr. Evil can be sure he's not the duplicate.  The second bet is fair by \emph{Indifference}.  Thus it turns out that accepting \textit{Indifference} in the original case where the principle was most plausible already leads to a diachronic Dutch book.  The defender of \emph{Indifference} finds the recommendations of the principle reasonable in cases like \textsc{Duplicating Dr. Evil} and thus must reject the idea that this kind of diachronic Dutch book demonstrates irrationality.  The Everettian Dutch book discussed above raises no new concerns for \emph{Indifference}.

The same point can be expressed in another way.  If the fission in \textsc{Once-or-Twice}, depicted in figure \ref{case1}, was classical fission instead of quantum fission then assigning credences in accord with \emph{Indifference} and betting accordingly would be reasonable (imagine cases of amoeba-like division as discussed in, e.g., \citealp{parfit1971}).  If these beliefs and betting behaviors are acceptable in the classical case, why not in the quantum case as well?  The charge of diachronic inconsistency does not raise distinctively quantum problems for \emph{Indifference}.

\section{Circularity}\label{subsys}

There are reasons to be concerned that an appeal to reduced density matrices in deriving the Born rule involves some sort of illegitimate circularity (\citealp{zeh1997,baker2007}; \citealp[\textsection 8.2.2]{schlosshauer}; \citeauthor{zurek2003b}, \citeyear{zurek2003b}, \citeyear{zurek2005}, \citeyear{zurek2010}; \citealp[\textsection 5.4]{kent2010}). This objection comes in a variety of forms.  We'll consider three.

First, one might be concerned that the very use of inner products, partial trace operations, and reduced density matrices is forbidden until one has derived a connection between squared inner product and probability.  This cannot be right.  How could one derive such a thing without ever writing an inner product?  Partial trace operations and reduced density matrices are perfectly well-defined mathematically within the framework of quantum theory.  Their definition does not \emph{require} understanding any number as a probability.  The concern about circularity must then be a concern about the \emph{physical interpretation} of these mathematical objects \citep[see also][FAQ \#6]{zurek2010}.

Second, the idea that the reduced density matrix describes the state of a subsystem---which we use in our derivation of the Born rule---could be doubted.
\begin{quotation}
\noindent\small{{It is not obvious that the reduced density operator for system $A$ is in any sense a description for the state of system $A$. The physical justification for making this identification is that the reduced density operator provides the correct measurement statistics for measurements made on system $A$.  \citep[\textsection 2.4.3]{nielsen2010}}}
\end{quotation}
It is true that the use of reduced density matrices to describe subsystems is often motivated by the fact that the reduced density matrix retains enough information about the state to deduce Born rule probabilities for outcomes of measurements performed solely on that system (\citealp[\textsection 2.4.6]{schlosshauer}; \citealp[box 2.6]{nielsen2010}).  However, we believe an alternative justification is possible which has nothing to do with probabilities or measurements.  (We only give a first pass at such a justification here.  In particular, we do not show that reduced density matrices are the only viable  way to mathematically represent the state of a subsystem.\footnote{Alternatively, one might appeal to Zurek's envariance-based argument for the use of reduced density matrices without accepting his later steps concerning probabilities \citep[II.B]{zurek2005}.})  In \textsection \ref{sec21} we proposed that the mathematical representation of the state of a subsystem should (1) together with the states of all other subsystems and facts about the connections between the subsystems yield the total state, and (2) be sufficient to determine its own evolution when the subsystem is isolated.  In general, the reduced density matrix for a composite system $AB$, $\widehat{\rho}_{AB}$, cannot be constructed uniquely from the separated reduced density matrices for $A$ and $B$.  The matrix $\widehat{\rho}_{AB}$ also encodes facts about the entanglement between $A$ and $B$.  But, as these are just facts about how $A$ and $B$ are connected, condition (1) is satisfied.  To see that condition (2) is met, suppose that at least for a time subsystem $A$ is isolated from everything else, $E$.  Let $\widehat{U}_t$ be the unitary operator that gives the time evolution of the total state.  Since $A$ and $E$ are non-interacting, $\widehat{U}_t=\widehat{U}_A\otimes\widehat{U}_E$.  The time evolution of $\widehat{\rho}_A$ is then given by $\widehat{U}_A\widehat{\rho}_A\widehat{U}_A^\dagger$ ($\widehat{U}_E$ is irrelevant).

Third, there is a worry that we cannot assume a structure of branching quantum worlds without assuming squared-amplitudes give probabilities.  Suppose that the reduced density matrix for a macroscopic system is approximately diagonal.  Why are we justified in treating the diagonal terms as branches and the off-diagonal terms as somehow unimportant and certainly not branches themselves?  (See \citealp{baker2007}; brief reply in \citealp[pp. 243-254]{wallace2012}.)  Our derivation treats the probabilities involved as purely \emph{epistemic}, quantifying an agent's uncertainty about which world they are in.  The question of how patterns in the wave function give rise to distinct worlds---and people who can wonder about which world they are in---is primarily \emph{metaphysical}.  In this article we've offered no additional insights on how this important project is to be completed, choosing to work under the assumption that it can be.

It should be noted that the use of density matrices, while convenient, is not strictly necessary for our derivation of the Born rule.  The proof of the Born rule could equally well be derived using purely the language of state vectors.  Instead of starting from the idea that the state of the environment is irrelevant, we could begin with the closely related thought that changes in the environment---represented by unitary transformations---ought not change the probabilities one assigns. In \citet{carroll2013} we gave a derivation along these lines.

\section{Generalization of the Born Rule Derivation}\label{generalization}

In the main text we showed how one can derive the Born rule in two simple cases.  Here we extend the arguments of \textsection \ref{prf} to the more general case of $N$ orthogonal branches with arbitrary rational squared-amplitudes and arbitrary phases.  As the proof here is restricted to \emph{rational} squared-amplitudes, one might be concerned about whether the account gives any advice at all when the squared-amplitudes are irrational.  This is especially worrisome as there are more irrational numbers than rational numbers.  However, if we assume (plausibly) that the probabilities vary continuously with small changes in the amplitudes, the restricted proof is sufficient. For any wave function with irrational squared-amplitudes there exist arbitrarily similar wave functions with rational squared-amplitudes (as the rationals are a dense subset of the reals).

A general state of the above form can be written as
\begin{align}
\ket{\Psi_0}&= \frac{c_1}{T}e^{i\theta_1}\ket{R}_A\ket{d_1}_D\ket{E_1}+\frac{c_2}{T}e^{i\theta_2}\ket{R}_A\ket{d_2}_D\ket{E_2}+...+\frac{c_N}{T}e^{i\theta_N}\ket{R}_A\ket{d_N}_D\ket{E_N}
\nonumber
\\
&= \sum_{k=1}^N{\frac{c_k}{T}e^{i\theta_k}\ket{R}_A\ket{d_k}_D\ket{e_k}_E}\ ,
\label{}
\end{align}
where each $c_k>0$, $c_k^2\in\mathbb{Z}^+$, $\theta_k\in\mathbb{R}$, and $T=\sqrt{\sum_k{c_k^2}}$.  $\ket{R}_A$ is Alice's state which is taken to be the same on every branch.  $\ket{d_k}_D$ are possible states of the detector or system of interest, $D$, the entity whose states Alice is assigning probabilities to.  $\ket{E_k}$ are the possible states of the environment.  Here Alice would like to know what probabilities to assign to each possible state $d_k$.  We will show that the appropriate probability of $d_k$ is $\frac{c_k^2}{T^2}$ (unless $d_k=d_j$ for some $j$, a situation where the particle is in the same state for two or more different environments, in which case the probability is given by the sum of the amplitude-squared for each $j$ such that $d_k=d_j$ including $j=k$, $\sum_j\frac{c_j^2}{T^2}$).  The probability that Alice ought to assign to each $p_k$ will be unchanged if we make the following transformation on each environment state (this follows from \emph{ESP-QM}, since the transformation leaves the Alice+Detector reduced density matrix unaffected),
\begin{equation}
\ket{E_k}\longrightarrow\frac{1}{c_k}\left\{|E_{k,1}'\rangle+...+|E_{k,c_k^2}'\rangle\right\}\ .
\label{}
\end{equation}
The environment state $E_k$ is taken to a superposition of $c_k^2$ different environment states.
These transformations take $\Psi_0$ to
\begin{equation}
\ket{\Psi_{eqamp}}=\sum_{k=1}^N{\sum_{j=1}^{c_k^2}\frac{1}{T}e^{i\theta_k}\ket{R}_A\ket{d_k}_D\ket{E_{k,j}'}}\ .
\label{eqamp}
\end{equation}
In this state, all of the components have equal amplitude.  Further, there are $c_k^2$ terms where $D$ is in state $d_k$ (unless $d_k=d_j$ for some $j$, in which case there are $\sum_j c_j^2$ terms).  Presently we will prove that each component of $\Psi_{eqamp}$ has equal probability.  Looking at the number of terms with each $d_k$, it is clear that this is all that is needed to show that the probabilities in $\Psi_0$ for each $d_k$ are given by the Born rule.  The above method for moving from a state of unequal amplitudes to one of equal amplitudes is also used in \citep[\textsection II.D]{zurek2005}.

We've reduced the problem to showing that terms in equal amplitude superpositions are equiprobable.  A general state with equal amplitude terms, such as $\Psi_{eqamp}$, can be written as
\begin{equation}
\ket{\Psi}=\sum_{k=1}^N{\frac{1}{\sqrt{N}}e^{i\theta_k}\ket{R}_A\ket{d_k}_D\ket{E_k}}\ .
\label{psipsipsi}
\end{equation}
Here some states $d_k$ may be identical.  Now consider two possible transformations which leave the Alice+Detector reduced density matrix unchanged.  First, consider a transformation which entangles $N$ display screens with each environment state $E_k$.  The symbol on the first screen, symbol $1$, may be either $S_1$ or $S_1'$, maybe either $\heartsuit$ or $\diamondsuit$, symbol 2 might be $S_2$ or $S_2'$, etc.  The first transformation entangles the $k$th state with $N$ displays, most of which show the unprimed symbols but the $k$th display shows the primed symbol:
\begin{equation}
\ket{E_k}\longrightarrow\ket{S_1}\ket{S_2}...\ket{S_k'}...\ket{S_N}\ket{E_k^*}\ .
\label{}
\end{equation}
This takes the state $\Psi$ to
\begin{equation}
\ket{\Psi_\alpha}=\sum_{k=1}^N{\frac{1}{\sqrt{N}}e^{i\theta_k}\ket{R}_A\ket{d_k}_D\ket{S_1}...\ket{S_k'}...\ket{S_N}\ket{E_k^*}}\ .
\label{psialphaalpha}
\end{equation}
The second transformation gives most environment states the same set of symbols, except the $N$th state which gets all of the primed symbols,
\begin{equation}
\ket{E_k}\longrightarrow\left\{\begin{array}{ll} \ket{S_1}\ket{S_2}...\ket{S_N}\ket{E_k^{**}} & \text{   if }k\neq N \\\\ \ket{S_1'}\ket{S_2'}...\ket{S_N'}\ket{E_k^{**}} & \text{   if }k= N  \end{array}\right.\ .
\label{}
\end{equation}
This takes $\Psi$ to
\begin{align}
\ket{\Psi_\beta}&= \sum_{k=1}^{N-1}{\frac{1}{\sqrt{N}}e^{i\theta_k}\ket{R}_A\ket{d_k}_D\ket{S_1}\ket{S_2}...\ket{S_N}\ket{E_k^{**}}}
\nonumber
\\
&\qquad+ \frac{1}{\sqrt{N}}e^{i\theta_N}\ket{R}_A\ket{d_N}_D\ket{S_1'}\ket{S_2'}...\ket{S_N'}\ket{E_N^{**}}\ .
\label{}
\end{align}
Now we use similar techniques as in \textsection \ref{prf} to show that the probabilities of each component of $\Psi_\alpha$ are the same as the probability of the $N$-th component of $\Psi_\beta$.  Focusing on the reduced density matrix of Alice+Display $k$, we see that 
\begin{equation}
P\left(d_k\middle|\Psi_\alpha\right)=P\left(d_N\middle|\Psi_\beta\right)\ .
\end{equation}
Combining these results for all $k$, we see that the probability of each term in $\Psi_\alpha$ must be equal and thus that the probability of each $d_k$ in state $\Psi$ is given by the Born rule (because $\Psi_\alpha$, \eqref{psialphaalpha}, and $\Psi$, \eqref{psipsipsi}, agree on the Alice+Detector reduced density matrix).  In combination with the discussion in the previous paragraph, this completes the proof that the probabilities for different states $d_k$ in the general state $\Psi_0$ are given by the Born rule.\\

\hfill \textit{Charles T. Sebens}

\hfill \textit{Department of Philosophy, University of Michigan}

\hfill \textit{435 South State Street, Ann Arbor, MI 48109, USA}

\hfill \textit{csebens@gmail.com}\\

\hfill \textit{Sean M. Carroll}

\hfill \textit{Department of Physics, California Institute of Technology}

\hfill \textit{1200 East California Boulevard, Pasadena, CA 91125, USA}

\hfill \textit{seancarroll@gmail.com}

\bibliography{SLUOPbibliographyfile}

\begin{thebibliography}{}

\bibitem[\protect\citename{Aguirre \& Tegmark, }{[2011]}]{Aguirre:2010rw}
Aguirre, A., \& Tegmark, M. {[2011]}.
\newblock `Born in an Infinite Universe: A Cosmological Interpretation of
  Quantum Mechanics'.
\newblock {\em Physical Review D}, {\bf 84}, 105002.

\bibitem[\protect\citename{Albert, }{[2010]}]{albert2010}
Albert, D.~Z. {[2010]}.
\newblock `Probability in the Everett Picture'.
\newblock  pp. 355--68.
\newblock in \cite{MWbook}.

\bibitem[\protect\citename{Albrecht \& Phillips, }{[2014]}]{Albrecht:2012zp}
Albrecht, A., \& Phillips, D. {[2014]}.
\newblock {`Origin of Probabilities and their Application to the Multiverse'}.
\newblock {\em Physical Review D}, {\bf 90}, 123514.

\bibitem[\protect\citename{Arntzenius, }{[2003]}]{arntzenius2003}
Arntzenius, F. {[2003]}.
\newblock `Some Problems for Conditionalization and Reflection'.
\newblock {\em The Journal of Philosophy}, {\bf 100}(7), pp. 356--70.

\bibitem[\protect\citename{Baker, }{[2007]}]{baker2007}
Baker, D.~J. {[2007]}.
\newblock `Measurement Outcomes and Probability in Everettian Quantum
  Mechanics'.
\newblock {\em Studies In History and Philosophy of Science Part B: Studies In
  History and Philosophy of Modern Physics}, {\bf 38}(1), pp. 153--69.

\bibitem[\protect\citename{Bostrom, }{[2002]}]{bostrom2002}
Bostrom, N. {[2002]}.
\newblock {\em Anthropic Bias: Observation Selection Effects in Science and
  Philosophy}.
\newblock Routledge.

\bibitem[\protect\citename{Bradley, }{[2011]}]{bradley2011}
Bradley, D.~J. {[2011]}.
\newblock `Confirmation in a Branching World: The Everett Interpretation and
  Sleeping Beauty'.
\newblock {\em The British Journal for the Philosophy of Science}, {\bf 62}(2),
  pp. 323--42.

\bibitem[\protect\citename{Bradley, }{[forthcoming]}]{bradley2014}
Bradley, D.~J. {[forthcoming]}.
\newblock `Everettian Confirmation and Sleeping Beauty: Reply to Wilson'.
\newblock {\em The British Journal for the Philosophy of Science}.

\bibitem[\protect\citename{Carroll \& Sebens, }{[2014]}]{carroll2013}
Carroll, S.~M., \& Sebens, C.~T. {[2014]}.
\newblock `Self-Locating Uncertainty and the Origin of Probability in
  Everettian Quantum Mechanics'.
\newblock {\em Pages  157--69 of:} Struppa, D., \& Tollaksen, J. (eds), {\em
  Quantum Theory: A Two-Time Success Story, Yakir Aharonov Festschrift}.
\newblock Springer.

\bibitem[\protect\citename{Deutsch, }{[1999]}]{deutsch1999}
Deutsch, D. {[1999]}.
\newblock `Quantum Theory of Probability and Decisions'.
\newblock {\em Proceedings of the Royal Society of London}, {\bf A458}, pp.
  3129--37.

\bibitem[\protect\citename{Dizadji-Bahmani, }{[forthcoming]}]{dizadji2013}
Dizadji-Bahmani, F. {[forthcoming]}.
\newblock `The Probability Problem in Everettian Quantum Mechanics Persists'.
\newblock {\em The British Journal for the Philosophy of Science}.

\bibitem[\protect\citename{Elga, }{[2004]}]{elga2004}
Elga, A. {[2004]}.
\newblock `Defeating Dr. Evil with Self-Locating Belief'.
\newblock {\em Philosophy and Phenomenological Research}, {\bf 69}(2), pp.
  383--96.

\bibitem[\protect\citename{Freivogel, }{[2011]}]{Freivogel:2011eg}
Freivogel, B. {[2011]}.
\newblock {`Making Predictions in the Multiverse'}.
\newblock {\em Classical and Quantum Gravity}, {\bf 28}, 204007.

\bibitem[\protect\citename{Gleason, }{[1957]}]{gleason}
Gleason, A.~M. {[1957]}.
\newblock `Measures on the Closed Subspaces of a Hilbert Space'.
\newblock {\em Journal of Mathematics and Mechanics}, {\bf 6}, pp. 885--94.

\bibitem[\protect\citename{Greaves, }{[2004]}]{greaves2004}
Greaves, H. {[2004]}.
\newblock `Understanding Deutsch's Probability in a Deterministic Multiverse'.
\newblock {\em Studies In History and Philosophy of Science Part B: Studies In
  History and Philosophy of Modern Physics}, {\bf 35}(3), pp. 423--56.

\bibitem[\protect\citename{Greaves, }{[2007a]}]{greaves2007b}
Greaves, H. {[2007a]}.
\newblock `On the Everettian Epistemic Problem'.
\newblock {\em Studies In History and Philosophy of Science Part B: Studies In
  History and Philosophy of Modern Physics}, {\bf 38}(1), pp. 120--52.

\bibitem[\protect\citename{Greaves, }{[2007b]}]{greaves2007}
Greaves, H. {[2007b]}.
\newblock `Probability in the Everett Interpretation'.
\newblock {\em Philosophy Compass}, {\bf 2}(1), pp. 109--28.

\bibitem[\protect\citename{Greaves \& Myrvold, }{[2010]}]{greavesM}
Greaves, H., \& Myrvold, W. {[2010]}.
\newblock `Everett and Evidence'.
\newblock  pp. 264--304.
\newblock in \cite{MWbook}.

\bibitem[\protect\citename{Groisman {\em et~al.\ }\relax,
  }{[2013]}]{groisman2013}
Groisman, B., Hallakoun, N., \& Vaidman, L. {[2013]}.
\newblock `The Measure of Existence of a Quantum World and the Sleeping Beauty
  Problem'.
\newblock {\em Analysis}, {\bf 73}(4), pp. 695--706.

\bibitem[\protect\citename{Hartle \& Srednicki, }{[2007]}]{hartle2007}
Hartle, J.~B., \& Srednicki, M. {[2007]}.
\newblock `Are We Typical?'.
\newblock {\em Physical Review D}, {\bf 75}(12), 123523.

\bibitem[\protect\citename{Kent, }{[2010]}]{kent2010}
Kent, A. {[2010]}.
\newblock `One World Versus Many: The Inadequacy of Everettian Accounts of
  Evolution, Probability, and Scientific Confirmation'.
\newblock  pp. 307--54.
\newblock in \cite{MWbook}.

\bibitem[\protect\citename{Kent, }{[2014]}]{kent2014}
Kent, A. {[2014]}.
\newblock `Does it Make Sense to Speak of Self-Locating Uncertainty in the
  Universal Wave Function? Remarks on Sebens and Carroll'.
\newblock {\em Foundations of Physics},  1--7.

\bibitem[\protect\citename{Lewis, }{[1979]}]{lewis1979}
Lewis, D.~K. {[1979]}.
\newblock `Attitudes \textit{De Dicto} and \textit{De Se}'.
\newblock {\em The Philosophical Review}, {\bf 88}(4), pp. 513--43.

\bibitem[\protect\citename{Lewis, }{[2007]}]{plewis}
Lewis, P.~J. {[2007]}.
\newblock `Quantum Sleeping Beauty'.
\newblock {\em Analysis}, {\bf 67}(293), pp. 59--65.

\bibitem[\protect\citename{Lewis, }{[2009a]}]{lewis2009}
Lewis, P.~J. {[2009a]}.
\newblock `Probability, Self-Location, and Quantum Branching'.
\newblock {\em Philosophy of Science}, {\bf 76}(5), pp. 1009--19.

\bibitem[\protect\citename{Lewis, }{[2009b]}]{lewis2009reply}
Lewis, P.~J. {[2009b]}.
\newblock `Reply to Papineau and Dur{\`a}-Vil{\`a}'.
\newblock {\em Analysis}, {\bf 69}(1), pp. 86--9.

\bibitem[\protect\citename{Manley, }{[unpublished]}]{ManleyF}
Manley, D. {[unpublished]}.
\newblock `On Being a Random Sample'.

\bibitem[\protect\citename{Maudlin, }{[2014]}]{maudlinreview}
Maudlin, T. {[2014]}.
\newblock `\textit{Critical Study} David Wallace, \textit{The Emergent
  Multiverse: Quantum Theory According to the Everett Interpretation}'.
\newblock {\em No\^{u}s}, {\bf 48}(4), pp. 794--808.

\bibitem[\protect\citename{Meacham, }{[2008]}]{meacham2008}
Meacham, C. {[2008]}.
\newblock `Sleeping Beauty and the Dynamics of De Se Beliefs'.
\newblock {\em Philosophical Studies}, {\bf 138}(2), pp. 245--69.

\bibitem[\protect\citename{Nielsen \& Chuang, }{[2010]}]{nielsen2010}
Nielsen, M.~A., \& Chuang, I.~L. {[2010]}.
\newblock {\em Quantum Computation and Quantum Information}.
\newblock Cambridge University Press.

\bibitem[\protect\citename{Page, }{[2007]}]{page2007}
Page, D.~N. {[2007]}.
\newblock `Typicality Defended'.
\newblock $\langle$arxiv.org/abs/0707.4169$\rangle$.

\bibitem[\protect\citename{Page, }{[2009a]}]{Page:2009qe}
Page, D.~N. {[2009a]}.
\newblock {`The Born Rule Fails in Cosmology'}.
\newblock {\em Journal of Cosmology and Astroparticle Physics}, {\bf 0907},
  008.

\bibitem[\protect\citename{Page, }{[2009b]}]{Page:2009mb}
Page, D.~N. {[2009b]}.
\newblock {`Born Again'}.
\newblock $\langle$arxiv.org/abs/0907.4152$\rangle$.

\bibitem[\protect\citename{Page, }{[2010]}]{Page:2010bj}
Page, D.~N. {[2010]}.
\newblock {`Born's Rule Is Insufficient in a Large Universe'}.
\newblock $\langle$arxiv.org/abs/1003.2419$\rangle$.

\bibitem[\protect\citename{Papineau, }{[1996]}]{papineau}
Papineau, D. {[1996]}.
\newblock `Many Minds are No Worse than One'.
\newblock {\em The British Journal for the Philosophy of Science}, {\bf 47}(2),
  pp. 233--41.

\bibitem[\protect\citename{Papineau \& Dur{\`a}-Vil{\`a},
  }{[2009a]}]{papineau2009}
Papineau, D., \& Dur{\`a}-Vil{\`a}, V. {[2009a]}.
\newblock `{\mockalph{bb}}A Thirder and an Everettian: A Reply to Lewis's
  ``Quantum Sleeping Beauty'''.
\newblock {\em Analysis}, {\bf 69}(1), pp. 78--86.

\bibitem[\protect\citename{Papineau \& Dur{\`a}-Vil{\`a},
  }{[2009b]}]{papineau2009reply}
Papineau, D., \& Dur{\`a}-Vil{\`a}, V. {[2009b]}.
\newblock `{\mockalph{dd}}Reply to Lewis: Metaphysics Versus Epistemology'.
\newblock {\em Analysis}, {\bf 69}(1), pp. 89--91.

\bibitem[\protect\citename{Parfit, }{[1971]}]{parfit1971}
Parfit, D. {[1971]}.
\newblock `Personal Identity'.
\newblock {\em The Philosophical Review}, {\bf 80}(1), pp. 3--27.

\bibitem[\protect\citename{Peterson, }{[2011]}]{peterson}
Peterson, D. {[2011]}.
\newblock `Qeauty and the Books: A Response to Lewis's Quantum Sleeping Beauty
  Problem'.
\newblock {\em Synthese}, {\bf 181}(3), pp. 367--74.

\bibitem[\protect\citename{Price, }{[2010]}]{price2010}
Price, H. {[2010]}.
\newblock `Decisions, Decisions, Decisions: Can Savage Salvage Everettian
  Probability?'.
\newblock  pp. 369--90.
\newblock in \cite{MWbook}.

\bibitem[\protect\citename{Salem, }{[2012]}]{Salem:2011qz}
Salem, M.~P. {[2012]}.
\newblock {`Bubble Collisions and Measures of the Multiverse'}.
\newblock {\em Journal of Cosmology and Astroparticle Physics}, {\bf 1201},
  021.

\bibitem[\protect\citename{Saunders, }{[2010a]}]{saunders2010}
Saunders, S. {[2010a]}.
\newblock `Many Worlds? An Introduction'.
\newblock  pp. 1--49.
\newblock in \cite{MWbook}.

\bibitem[\protect\citename{Saunders, }{[2010b]}]{saunders2010b}
Saunders, S. {[2010b]}.
\newblock `Chance in the Everett Interpretation'.
\newblock  pp. 181--205.
\newblock in \cite{MWbook}.

\bibitem[\protect\citename{Saunders \& Wallace, }{[2008]}]{saunders2008}
Saunders, S., \& Wallace, D. {[2008]}.
\newblock `Branching and Uncertainty'.
\newblock {\em The British Journal for the Philosophy of Science}, {\bf 59}(3),
  pp. 293--305.

\bibitem[\protect\citename{Saunders {\em et~al.\ }\relax, }{[2010]}]{MWbook}
Saunders, S., Barrett, J., Kent, A., \& Wallace, D. (eds). {[2010]}.
\newblock {\em Many Worlds?: Everett, Quantum Theory, \& Reality}.
\newblock Oxford University Press.

\bibitem[\protect\citename{Schlosshauer,
  }{[2005]}]{schlosshauer2005decoherence}
Schlosshauer, M.~A. {[2005]}.
\newblock `Decoherence, the Measurement Problem, and Interpretations of Quantum
  Mechanics'.
\newblock {\em Reviews of Modern Physics}, {\bf 76}(4), 1267.

\bibitem[\protect\citename{Schlosshauer, }{[2007]}]{schlosshauer}
Schlosshauer, M.~A. {[2007]}.
\newblock {\em `Decoherence and the Quantum-to-Classical Transition'}.
\newblock Springer Verlag.

\bibitem[\protect\citename{Schlosshauer \& Fine, }{[2005]}]{schlosshauer2005}
Schlosshauer, M.~A., \& Fine, A. {[2005]}.
\newblock `On Zurek's Derivation of the Born Rule'.
\newblock {\em Foundations of Physics}, {\bf 35}(2), pp. 197--213.

\bibitem[\protect\citename{Schwarz, }{[forthcoming]}]{schwarz}
Schwarz, W. {[forthcoming]}.
\newblock `Belief Update across Fission'.
\newblock {\em The British Journal for the Philosophy of Science}.

\bibitem[\protect\citename{Srednicki \& Hartle, }{[2010]}]{srednicki2010}
Srednicki, M., \& Hartle, J. {[2010]}.
\newblock `Science in a Very Large Universe'.
\newblock {\em Physical Review D}, {\bf 81}(12), 123524.

\bibitem[\protect\citename{Tappenden, }{[2011]}]{tappenden2011}
Tappenden, P. {[2011]}.
\newblock `Evidence and Uncertainty in Everett's Multiverse'.
\newblock {\em The British Journal for the Philosophy of Science}, {\bf 62}(1),
  pp. 99--123.

\bibitem[\protect\citename{Titelbaum, }{[2008]}]{titelbaum2008}
Titelbaum, M.~G. {[2008]}.
\newblock `The Relevance of Self-Locating Beliefs'.
\newblock {\em Philosophical Review}, {\bf 117}(4), pp. 555--606.

\bibitem[\protect\citename{Vaidman, }{[1998]}]{vaidman1998}
Vaidman, L. {[1998]}.
\newblock `On Schizophrenic Experiences of the Neutron or Why We Should Believe
  in the Many-Worlds Interpretation of Quantum Theory'.
\newblock {\em International Studies in the Philosophy of Science}, {\bf
  12}(3), pp. 245--61.

\bibitem[\protect\citename{Vaidman, }{[2011]}]{vaidman2011}
Vaidman, L. {[2011]}.
\newblock `Probability in the Many-Worlds Interpretation of Quantum Mechanics'.
\newblock {\em In:} Ben-Menahem, Y., \& Hemmo, M. (eds), {\em The Probable and
  the Improbable: Understanding Probability in Physics, Essays in Memory of
  Itamar Pitowsky}.
\newblock Springer.

\bibitem[\protect\citename{Vaidman, }{[2014a]}]{vaidman2014}
Vaidman, L. {[2014a]}.
\newblock Quantum Theory and Determinism.
\newblock {\em Quantum Studies: Mathematics and Foundations}, {\bf 1}, 5--38.

\bibitem[\protect\citename{Vaidman, }{[2014b]}]{vaidmanSEP}
Vaidman, L. {[2014b]}.
\newblock `Many-Worlds Interpretation of Quantum Mechanics'.
\newblock {\em In:} Zalta, E.~N. (ed), {\em The Stanford Encyclopedia of
  Philosophy}.
\newblock
  $\langle$plato.stanford.edu/archives/win2014/entries/qm-manyworlds/$\rangle$.

\bibitem[\protect\citename{Vaidman, }{[forthcoming]}]{vaidman2015}
Vaidman, L. {[forthcoming]}.
\newblock `Bell Inequality and Many-Worlds Interpretation'.
\newblock {\em In:} Bell, M., \& Gao, S. (eds), {\em Quantum Non-locality and
  Reality: 50 Years of Bell's Theorem}.
\newblock Cambridge University Press.

\bibitem[\protect\citename{Vilenkin, }{[2012]}]{Vilenkin:2013ik}
Vilenkin, A. {[2012]}.
\newblock {`Global Structure of the Multiverse and the Measure Problem'}.
\newblock {\em AIP Conference Proceedings}, {\bf 1514}, pp. 7--13.

\bibitem[\protect\citename{Wallace, }{[2002]}]{wallace2002}
Wallace, D. {[2002]}.
\newblock `Quantum Probability and Decision Theory, Revisited'.
\newblock $\langle$arxiv.org/abs/quant-ph/0211104$\rangle$.

\bibitem[\protect\citename{Wallace, }{[2003a]}]{wallace2003}
Wallace, D. {[2003a]}.
\newblock `Everett and Structure'.
\newblock {\em Studies In History and Philosophy of Science Part B: Studies In
  History and Philosophy of Modern Physics}, {\bf 34}(1), pp. 87--105.

\bibitem[\protect\citename{Wallace, }{[2003b]}]{wallace2003b}
Wallace, D. {[2003b]}.
\newblock `Everettian Rationality: Defending Deutsch's Approach to Probability
  in the Everett Interpretation'.
\newblock {\em Studies In History and Philosophy of Science Part B: Studies In
  History and Philosophy of Modern Physics}, {\bf 34}(3), pp. 415--39.

\bibitem[\protect\citename{Wallace, }{[2006]}]{wallace2006}
Wallace, D. {[2006]}.
\newblock `Epistemology Quantized: circumstances in which we should come to
  believe in the Everett interpretation'.
\newblock {\em The British Journal for the Philosophy of Science}, {\bf 57}(4),
  pp. 655--89.

\bibitem[\protect\citename{Wallace, }{[2007]}]{wallace2007}
Wallace, D. {[2007]}.
\newblock `Quantum Probability from Subjective Likelihood: Improving on
  Deutsch's Proof of the Probability Rule'.
\newblock {\em Studies In History and Philosophy of Science Part B: Studies In
  History and Philosophy of Modern Physics}, {\bf 38}(2), pp. 311--32.

\bibitem[\protect\citename{Wallace, }{[2010a]}]{wallace2010c}
Wallace, D. {[2010a]}.
\newblock `Decoherence and Ontology'.
\newblock  pp. 53--72.
\newblock in \cite{MWbook}.

\bibitem[\protect\citename{Wallace, }{[2010b]}]{wallace2010xi}
Wallace, D. {[2010b]}.
\newblock `Diachronic Rationality and Prediction-Based Games'.
\newblock {\bf CX}(3), pp. 243--66.

\bibitem[\protect\citename{Wallace, }{[2010c]}]{wallace2010b}
Wallace, D. {[2010c]}.
\newblock `How to Prove the Born Rule'.
\newblock  pp. 227--63.
\newblock in \cite{MWbook}.

\bibitem[\protect\citename{Wallace, }{[2012]}]{wallace2012}
Wallace, D. {[2012]}.
\newblock {\em The Emergent Multiverse: Quantum Theory According to the Everett
  Interpretation}.
\newblock Oxford University Press.

\bibitem[\protect\citename{Wallace, }{[2013]}]{wallace2013}
Wallace, D. {[2013]}.
\newblock `The Everett Interpretation'.
\newblock {\em Pages  460--88 of:} Batterman, R. (ed), {\em The Oxford Handbook
  of Philosophy of Physics}.

\bibitem[\protect\citename{Wilson, }{[2012]}]{wilson2012b}
Wilson, A. {[2012]}.
\newblock `Everettian Quantum Mechanics without Branching Time'.
\newblock {\em Synthese}, {\bf 188}(1), pp. 67--84.

\bibitem[\protect\citename{Wilson, }{[2013]}]{wilson2013objective}
Wilson, A. {[2013]}.
\newblock `Objective Probability in Everettian Quantum Mechanics'.
\newblock {\em The British Journal for the Philosophy of Science}, {\bf 64}(4),
  pp. 709--37.

\bibitem[\protect\citename{Wilson, }{[2014]}]{wilson2013}
Wilson, A. {[2014]}.
\newblock `Everettian Confirmation and Sleeping Beauty'.
\newblock {\em The British Journal for the Philosophy of Science}, {\bf 65}(3),
  pp. 573--98.

\bibitem[\protect\citename{Wilson, }{[forthcoming]}]{wilson2015}
Wilson, A. {[forthcoming]}.
\newblock `The Quantum Doomsday Argument'.
\newblock {\em The British Journal for the Philosophy of Science}.

\bibitem[\protect\citename{Zeh, }{[1997]}]{zeh1997}
Zeh, H. {[1997]}.
\newblock `What is Achieved by Decoherence?'.
\newblock {\em Pages  441--51 of:} Ferrero, M., \& van~der Merwe, A. (eds),
  {\em New Developments on Fundamental Problems in Quantum Physics}.
\newblock Springer.

\bibitem[\protect\citename{Zurek, }{[1981]}]{Zurek:1981xq}
Zurek, W.~H. {[1981]}.
\newblock `Pointer Basis of Quantum Apparatus: Into What Mixture Does the Wave
  Packet Collapse?'.
\newblock {\em Physical Review D}, {\bf 24}, pp. 1516--25.

\bibitem[\protect\citename{Zurek, }{[1982]}]{Zurek:1982ii}
Zurek, W.~H. {[1982]}.
\newblock `Environment Induced Superselection Rules'.
\newblock {\em Physical Review D}, {\bf 26}, pp. 1862--80.

\bibitem[\protect\citename{Zurek, }{[1993]}]{Zurek:1993ptp}
Zurek, W.~H. {[1993]}.
\newblock `Preferred States, Predictability, Classicality and the
  Environment-Induced Decoherence'.
\newblock {\em Progress of Theoretical Physics}, {\bf 89}(2), pp. 281--312.

\bibitem[\protect\citename{Zurek, }{[2003a]}]{zurek2003}
Zurek, W.~H. {[2003a]}.
\newblock `Decoherence, Einselection, and the Quantum Origins of the
  Classical'.
\newblock {\em Reviews of Modern Physics}, {\bf 75}(3), pp. 715--75.

\bibitem[\protect\citename{Zurek, }{[2003b]}]{zurek2003b}
Zurek, W.~H. {[2003b]}.
\newblock `Environment-Assisted Invariance, Entanglement, and Probabilities in
  Quantum Physics'.
\newblock {\em Physical Review Letters}, {\bf 90}(12), 120404.

\bibitem[\protect\citename{Zurek, }{[2005]}]{zurek2005}
Zurek, W.~H. {[2005]}.
\newblock `Probabilities from Entanglement, Born's Rule $p_ k=\left| \psi_
  {k}\right|^2$ from Envariance'.
\newblock {\em Physical Review A}, {\bf 71}(5), 052105.

\bibitem[\protect\citename{Zurek, }{[2010]}]{zurek2010}
Zurek, W.~H. {[2010]}.
\newblock `Quantum Jumps, Born's Rule, and Objective Reality'.
\newblock  pp. 409--432.
\newblock in \cite{MWbook}.

\end{thebibliography}

\end{document}